\documentclass[aps, prd, letterpaper, twocolumn, amsmath, amssymb, superscriptaddress, floatfix, nofootinbib, longbibliography, preprintnumbers, 10pt]{revtex4-2}

\usepackage{orcidlink}
\usepackage{graphicx, color}
\usepackage{booktabs}
\usepackage[letterspace=-10]{microtype} 

\usepackage{bm, amsmath, amsfonts, amssymb,xfrac}
\usepackage{multirow, tabularx, dcolumn, soul}
\usepackage{mathtools, leftidx, braket, slashed, cancel, bigdelim}
\usepackage{blkarray}
\usepackage[figures]{rotating}

\usepackage[utf8]{inputenc} 
\usepackage{hyperref}
\hypersetup{
    colorlinks=true,
    linkcolor=red,
    citecolor=blue,
    urlcolor=blue
}
\usepackage{pgf}
\usepackage{tikz}
\usepackage{lmodern}
\usepackage{fix-cm}

\newcommand\bef{\begin{figure}}
\newcommand\eef[1]{\label{fg:#1}\end{figure}}
\newcommand\beq{\begin{equation}}
\newcommand\eeq[1]{\label{#1}\end{equation}}
\newcommand\beqa{\begin{eqnarray}}

\newcommand\bet{\begin{table}}
\newcommand\eet[1]{\label{tb:#1}\end{table}}

\newcommand\cf{{\it c.f.}}

\def\IMSc{The Institute of Mathematical Sciences, CIT Campus, Chennai, 600113, India}
\def\HBNI{Homi Bhabha National Institute, Training School Complex, Anushaktinagar, Mumbai 400094, India}
\def\TIFR{Department of Theoretical Physics, Tata Institute of Fundamental Research, \\ Homi Bhabha Road, Mumbai 400005, India}
\newcommand{\imsc}{\affiliation{\IMSc}}
\newcommand{\hbni}{\affiliation{\HBNI}}
\newcommand{\tifr}{\affiliation{\TIFR}}
\begin{document}

\title{Lattice QCD Study of Positive Parity Dibaryons with Maximal Charm and Strangeness}

\author{Navdeep Singh Dhindsa~\orcidlink{0000-0002-3133-6979}}
\email{navdeep.s.dhindsa@gmail.com}
\imsc

\author{Nilmani Mathur~\orcidlink{0000-0003-2422-7317}}
\email{nilmani@theory.tifr.res.in}
\tifr

\author{M. Padmanath~\orcidlink{0000-0001-6877-7578}}
\email{padmanath@imsc.res.in}
\imsc
\hbni

\preprint{TIFR/TH/25-14}

\begin{abstract}
We present the ground-state energy spectra of dibaryons composed of single-flavor quarks, specifically systems with strangeness $\mathcal{S} = -6$ and charm $\mathcal{C} = 6$. Our lattice QCD study is based on $N_f=2+1+1$ MILC ensembles with highly improved staggered quark (HISQ) sea quarks, spanning four lattice spacings and two spatial volumes. We employ valence quark propagators realized using a relativistic overlap action, evaluate correlation matrices with carefully designed operator bases, and extract reliable ground-state energy estimates in the $S = 0$ and $S = 2$ spin channels. We explore their binding characteristics and interaction dynamics by examining the energy separation between the dibaryon states and the corresponding two-baryon thresholds. These results contribute to a deeper understanding of single-flavor dibaryon states as a function of the quark masses. In the $S=0$ channel, the $\Omega_{ccc}$-$\Omega_{ccc}$ system exhibits a clear signal of a bound state, while the $\Omega$-$\Omega$ system lies very close to the threshold, making it difficult to draw definitive conclusions. For $S=2$, both systems are found to be unbound.
\end{abstract}

\maketitle

\section{Introduction}\label{sec:intro}
Dibaryons form the simplest yet important subgroup in the hadron family that provides a well-defined framework for studying baryon-baryon interactions, which are key to our understanding of nuclear forces and binding in nuclei. Deuteron, a bound state of the $np$ system in coupled $^3S_1-^3D_1$ partial waves, remains to be the only stable dibaryon discovered, despite several decades-long extensive experimental searches. Recent investigations by the WASA-at-COSY collaboration have reported possible evidence for the existence of a light dibaryon resonance, $d^{*}(2380)$ \cite{WASA-at-COSY:2014dmv}, that is probably constituted of a $\Delta\Delta$ system bound by approximately 100 MeV, as predicted by Dyson and Xuong \cite{Dyson:1964xwa}. A comprehensive review of the past and the ongoing experimental efforts on various dibaryon channels and related theoretical proposals can be found in Ref. \cite{Clement:2016vnl}. 

The abundance of nucleon($N$)-rich targets and beams in experiments naturally implies the availability of a rich collection of $NN$ scattering data. However, dibaryons with strangeness have been relatively less accessible in experiments, owing to their large masses and relatively short lifetimes. On the theoretical front, there have been substantial efforts in understanding interactions in light dibaryons as well as dibaryons with strangeness. Investigations aiming at a deeper understanding of how hyperons interact with nucleons, with other hadrons, and between themselves can have important multidisciplinary implications. These studies provide crucial inputs in constraining Low Energy Constants (LEC) in Effective Field Theories (EFT) in describing nuclear many-body calculations relevant in experimental efforts using large nuclei reactions \cite{Epelbaum:2005pn}. They also can have important implications on the properties of astrophysical objects \cite{Bazavov:2014xya,Chatterjee:2015pua,Vidana:2017qey,Tolos:2020aln}. In this context, dibaryons with heavier quarks provide a valuable setting to study baryon-baryon interactions with minimal interference from chiral effects, offering a cleaner probe into the fundamental dynamics of strong interactions.

Earliest theoretical predictions for dibaryon states date back to 1963, when Oakes considered baryon-baryon states including the deuteron to form a multiplet, and predicted the existence and masses of other members in the multiplet \cite{Oakes:1963zza}. In 1964, Dyson and Xuong \cite{Dyson:1964xwa}, using the $SU(6)$ symmetry of color and light flavors, argued several light dibaryon states denoted as $D_{IJ}$, where $I$ and $J$ refer to the total isospin and total angular momentum. This included $D_{03}$, whose predicted mass was remarkably close to that of the observed $d^{*}(2380)$. Since $D_{03}$ ($d^*(2380)$) and $D_{30}$ belong to the same dibaryon multiplet, the experimental observation by the WASA-at-COSY collaboration of $d^*$ suggests the possible existence of the $D_{30}$ state. Unlike $d^*$, $D_{30}$ could be a truly exotic dibaryon, which can be composed entirely of six up or six down quarks, making its experimental search particularly compelling \cite{Clement:2016vnl}. By studying dibaryons composed entirely of strange quarks or their counterparts in the heavy quark sector, one can systematically approach calculations at lower quark masses, ultimately gaining insights into the formation of dibaryons in the light sector without strangeness, as originally predicted in Ref. \cite{Dyson:1964xwa}. There have been ongoing searches and theoretical proposals to explore isospin-3 dibaryons, such as those composed entirely of six up or six down quarks and analogous, fully strange dibaryons \cite{Bashkanov:2013cla, WASA-at-COSY:2016bha, Morita:2019rph, Bashkanov:2023yca}.

Theoretical studies on interactions with single-flavored two-baryon systems in the strange and heavy flavor sectors are limited. Two quantum channels that are relevant in the study of $S$-wave scattering of single-flavored two-baryon systems correspond to those with $J=$0 and 2. Different variants and extensions of quark potential models have been utilized to investigate these systems with light, strange, as well as heavy quark flavors \cite{Li:2000cb,Wang:1995bg,Huang:2013nba,Huang:2019hmq,Huang:2020bmb,Zhang:2000sv,Liu:2021pdu}. The predictions for scalar dibaryons vary from deeply bound with respect to the lowest strong decay thresholds, to shallow bound states, and to strong interaction unstable resonances, even for the heavy flavor sector. Although a general agreement on an unbound $J$=2 dibaryon system can be observed across various studies, the possibility of a bound state was not ruled out in some of the studies. Theoretical predictions for the binding in these systems remain conflicting, with different models yielding varying conclusions, notably the discrepancies between the Quark Delocalization Color Screening Model and the chiral quark model \cite{Huang:2019hmq}. 

In this context, lattice QCD methodologies provide a robust first-principles framework for investigating such dibaryon states while circumventing model-dependent assumptions and pave the way for a more definitive understanding of the interactions in these systems. However, lattice QCD studies of baryon-baryon scatterings and dibaryons face significant challenges, primarily due to the severe signal-to-noise problem in lattice QCD correlation functions \cite{Lepage:1989hd}, which worsens at large Euclidean time separations and complicates the reliable extraction of ground-state energies. This issue is particularly pronounced in light and strange dibaryons. As a result, lattice investigations in these sectors have often been performed at heavy and unphysical pion masses to mitigate signal-to-noise limitations. Recent progress in lattice QCD studies of dibaryons has been made across lighter and strange sectors \cf ~\cite{Wagman:2017tmp, Francis:2018qch, Iritani:2018vfn, Amarasinghe:2021lqa, Green:2021qol, Aoki:2023qih}. Additionally, significant advancements have been achieved in mitigating the signal-to-noise ratio problem in the investigations of multi-baryon systems \cite{Wagman:2024rid, Hackett:2024xnx, Chakraborty:2024scw, Ostmeyer:2024qgu}.

Two-baryon states involving only $\Omega$ or $\Omega_{ccc}$, referred to as $\mathcal{D}_{6s}$ and $\mathcal{D}_{6c}$ respectively, are appealing from a lattice QCD perspective as the interactions within are anticipated to be dominated by the valence strange or charm quark masses. The chiral light quark mass effects are only subdominant as the leading contributions are expected to arise from higher order two-pion exchange processes \cite{Tiburzi:2008bk}. This means the interactions in these systems rely less on extrapolation to the physical pion mass. The spin-3/2 baryons composed entirely of charm or strange quarks ($\Omega_{ccc}$ and $\Omega$) represent the lightest states in their respective sectors. Owing to their large masses, studies involving purely strange and charm quarks are computationally cheaper and signal-wise cleaner than their light flavor counterparts. For these reasons, the $\Omega$ baryon mass is often used as a benchmark for scale setting, owing to its clean signal, low computational cost, and precise experimental determination. Meanwhile, the $\Omega_{ccc}$ baryon serves as an excellent system for studying quark confinement dynamics, as it is largely unaffected by leading light-quark effects. Such dibaryons provide a valuable probe into the dynamics of two-baryon interactions at heavy quark masses, offering insights into whether these interactions lead to bound-state formation. The existence of any deeply bound state would require an explanation beyond the traditional meson-exchange mechanisms used in nuclear physics, as the meson masses would be very heavy.

Lattice QCD studies of singly flavored dibaryon systems are limited to a few, with only three lattice investigations available for the $\Omega$-$\Omega$ system \cite{Buchoff:2012ja,HALQCD:2015qmg,Gongyo:2017fjb} and a single study for $\Omega_{ccc}$-$\Omega_{ccc}$ \cite{Lyu:2021qsh} and $\Omega_{bbb}$-$\Omega_{bbb}$ \cite{Mathur:2022ovu}, respectively. In Ref. \cite{Buchoff:2012ja}, the authors investigated the $\Omega$-$\Omega$ system on $N_f = 2+1$ lattice QCD ensembles using anisotropic Wilson-clover fermions and a pion mass of $\sim400$ MeV. The finite-volume scattering analysis performed following L\"uscher's prescription \cite{Luscher:1990ux} in Ref.~\cite{Buchoff:2012ja} suggested a weakly repulsive interaction in both the $S=0$ and $S=2$ channels. In contrast, a HALQCD study with $N_f = 2+1$ PACS-CS configurations indicates an attractive interaction, though not strong enough to form a tightly bound state at $m_\pi \approx 700$ MeV \cite{HALQCD:2015qmg}. A follow-up investigation at a near-physical pion mass of $m_\pi \approx 146$ MeV, the interaction remains weakly attractive, with a potential strength sufficient to support a very shallow bound state in the $S=0$ channel \cite{Gongyo:2017fjb}. On similar lines, the $\Omega_{ccc}$-$\Omega_{ccc}$ system in the $S=0$ channel on the same near-physical pion mass ensembles was observed to be strong enough to host a shallow bound state \cite{Lyu:2021qsh}, although inclusion of Coulombic interactions were found to make the system unbound. In our previous publication \cite{Mathur:2022ovu}, we found the $\Omega_{bbb}$-$\Omega_{bbb}$ system to be bound by around 80 MeV with respect to $2M_{\Omega_{bbb}}$. 

To gain a more comprehensive understanding of single-flavor dibaryons towards lighter quark masses, we investigate the $\Omega$-$\Omega$ and $\Omega_{ccc}$-$\Omega_{ccc}$ systems using a similar ensemble setup as utilized in our previous study \cite{Mathur:2022ovu}. Our study aims to gain a deeper understanding of
 their binding properties and interaction strengths in the $S$-wave interactions in these singly flavored dibaryon systems with $J=0$ and $J=2$ quantum numbers. 

The paper is organized as follows: Section \ref{sec:latt} outlines the lattice setup used in this work, detailing the valence and sea actions, the interpolators used, evaluation of correlation matrices, and the subsequent analysis of our lattice setup. The results are presented in Section \ref{sec:results}, followed by conclusions and discussion in Section \ref{sec:conc}. 

%%%%%%%%%%%%%%%%%%%%%%%%%%%%%%%%%%%%%%%%%%
\section{Lattice Details and Operator Construction}\label{sec:latt}

\begin{figure}[ht]
\includegraphics[width=0.45\textwidth]{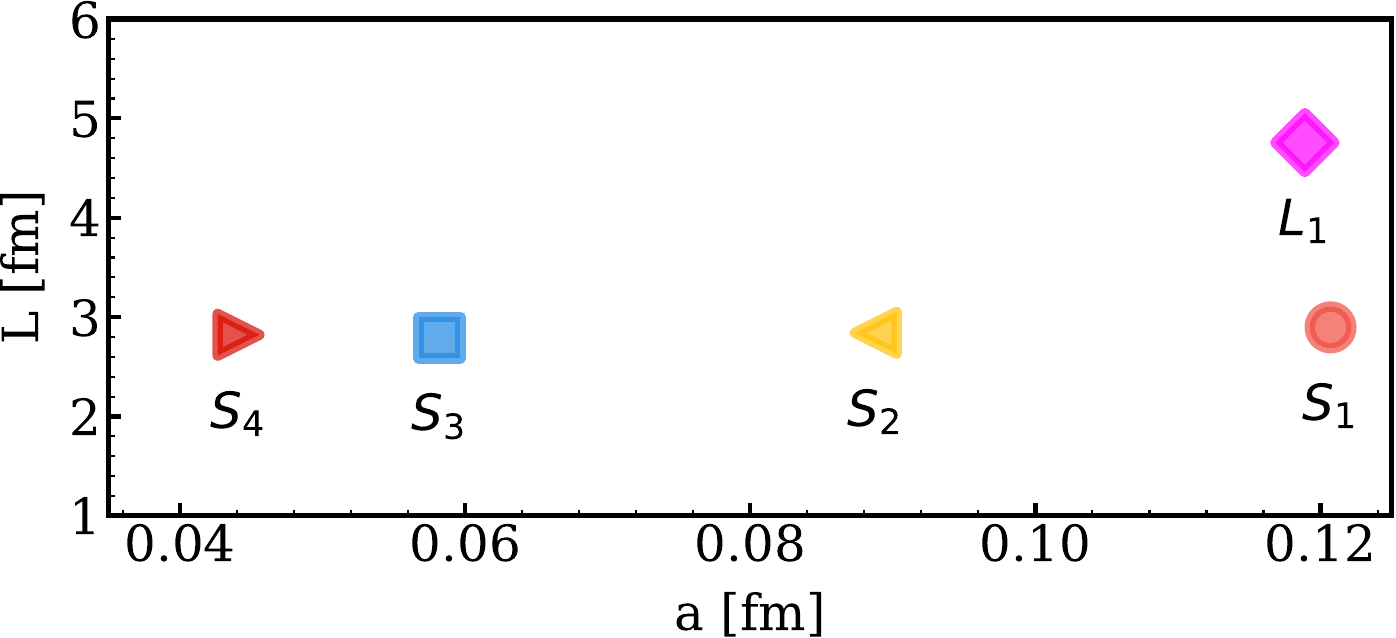}
\caption{\label{fig:lattice} Five lattice QCD ensembles are used in this work. The markers assigned to each ensemble will remain consistent throughout the paper. Refer to Table \ref{tab:pars} for detailed ensemble specifications.}
\end{figure}

We adopt the same lattice setup as used in our previous works \cite{Junnarkar:2018twb, Mathur:2018epb, Mathur:2018rwu, Junnarkar:2019equ, Junnarkar:2022yak, Mathur:2022ovu, Padmanath:2023rdu, Radhakrishnan:2024ihu, Dhindsa:2024erk, Junnarkar:2024kwd, Tripathy:2025vao}. For self-containment, we provide a brief discussion here. In this work, we utilize five lattice QCD ensembles featuring two distinct spatial volumes, four distinct lattice spacings, and $N_f = 2+1+1$ dynamical Highly Improved Staggered Quark (HISQ) fields, generated by the MILC collaboration \cite{MILC:2012znn, Bazavov:2017lyh}. We present the ensemble details in Fig. \ref{fig:lattice} and Table \ref{tab:pars}. The strange and charm quark masses in the sea sector are tuned to their physical values, while the light quarks are set heavier than their physical values, with the light-to-strange quark mass ratio ($m_l/m_s$) set as $\{0.1, 0.2\}$ (See the Table \ref{tab:pars} for details). The gauge field dynamics are governed by a one-loop, tadpole-improved Symanzik gauge action, with coefficients tuned through $\mathcal{O}(\alpha_s a^2, n_f\alpha_s a^2)$ \cite{Follana:2006rc}.

\begingroup
\renewcommand*{\arraystretch}{1.5}
\begin{table}[ht]
	\centering
	\begin{tabular}{cccccc} \hline \hline 
	Ens & $L^3 \times T$     &  $m_l/m_s$ & $m^{\text{sea}}_\pi$ (MeV) & $m_\pi \text{L}$ & $a$ (fm) \\ \hline \hline 
	$S_4$ & $64^3 \times 192 $ & 1/5 &             315           &  4.29      & 0.042 \\ \hline 
	$S_3$ & $48^3 \times 144 $ &     1/5 &         319           & 4.52       & 0.0582 \\ \hline   
    $S_2$ & $32^3 \times 96 $  &     1/5 &         313           & 4.51       & 0.0888 \\ \hline   
	$S_1$ & $24^3 \times 64 $  &     1/5 &         305           & 4.48       & 0.1207  \\ \hline
    $L_1$ & $40^3 \times 64 $  &     1/10 &         217           &  5.23      & 0.1189 \\ \hline \hline
	\end{tabular}
	\caption{\label{tab:pars}{Parameters of lattice QCD ensembles used in this work. More details about the ensembles can be found in the Refs.\cite{MILC:2012znn, Bazavov:2017lyh}.}}
	\end{table} 
\endgroup

In the valence sector, we use an overlap action to describe the dynamics of strange and charm quarks \cite{Neuberger:1997fp, Neuberger:1998wv}. The valence strange quark mass is tuned to its physical value by ensuring that the lattice-determined mass of the fictitious pseudoscalar $\bar{s}s$ meson matches 688.5 MeV, a value derived from pion and kaon masses in lattice simulations \cite{Chakraborty:2014aca}. The valence charm quark mass is tuned on each ensemble by setting the kinetic mass of the spin-averaged 1S charmonium state, $a\overline{M}_{\bar{c}c}^{\text{kin}} = 0.75 aM_{\text{kin}}(J/\psi) + 0.25 aM_{\text{kin}}(\eta_c)$, to the physical value of 3068 MeV, following the Fermilab prescription \cite{El-Khadra:1996wdx}. More details about this method of charm mass tuning can be found in our work \cite{Basak:2012py,Basak:2013oya}, where we also explored an alternative approach based on tuning the charm mass using the pole mass of mesons. 

\subsection{Interpolators}
The finite-volume energy spectra are extracted from two-point correlation functions given by:  
\begin{eqnarray}
C_{ij}(t_f-t_i) &=& \sum_{\mathbf{x}} \left\langle \Phi_i(\mathbf{x},t_f) \tilde\Phi_j^\dagger(t_i) \right\rangle \nonumber \\ &=& \sum_{n} \frac{Z_i^n \tilde{Z}_j^{n\dagger}}{2E^n} e^{-E^n (t_f-t_i)},
\label{eq:corr}
\end{eqnarray}  
where $\{\Phi_i(\mathbf{x}, t)\}$ are carefully chosen interpolating operators optimized for maximal overlap with the states of interest. The overlap factors $ Z_i^n = \langle 0 | \Phi_i | n \rangle $ encode the coupling of these operators to energy eigenstates. The exponential time dependence evident in the spectral decomposition provides access to the ground state energies at asymptotic times. 

A wall smearing of the quark-fields on Coulomb-gauge fixed configurations is assumed at the source time slice ($t_i$) to optimize overlap with low-momentum states. This setup effectively eliminates nonzero momentum modes in the free-field theory. This method is empirically observed to suppress excited-state contamination in interacting scenarios and has been utilized in Lattice QCD calculations for a long time. A point sink maintains a well-defined signal in the correlation measurements with minimal smearing effects, facilitating precise energy extraction and ensuring robust ground-state mass determination. However, the combination of wall sources and point sinks leads to an asymmetric setup, and thus results in non-Hermitian correlation matrices. To highlight this asymmetric nature of correlation matrices, we represent the operators at the source timeslice and the corresponding overlap factors with a $``~\tilde{}~"$ in Eq. \eqref{eq:corr}. This may result in negative and complex spectral amplitudes in Eq. \eqref{eq:corr}, potentially leading to incorrect estimation of the true ground state energy. As demonstrated in our earlier works \cite{Padmanath:2023rdu, Radhakrishnan:2024ihu, Tripathy:2025vao} on multi-quark systems, we employ not only the wall-source to point-sink setup but also alternative strategies for constructing correlation functions. These alternatives include the use of different operator choices and sink smearing techniques such as box smearing \cite{Hudspith:2020tdf}. These approaches, combined with careful fitting, ensure reliable ground-state energy extraction with excited-state contamination mitigated by appropriately chosen fit windows.

To enhance the access to ground state energy, we evaluate correlation matrices, $C(t)$, for a basis of interpolating operators, $\{\Phi_i(\mathbf{x}, t)\}_N$, where $N$ is the number of operators in the basis. $C(t)$ are solved variationally following the solutions of a generalized eigenvalue problem: $C(t) v_n = \lambda_n(t, t_0) C(t_0) v_n$. Here, the eigenvalue correlators $\lambda_n(t, t_0)$ carry all the time dependence information and exhibit an approximately single-exponential behavior at large Euclidean times, enabling a systematic extraction of ground state energy. The eigenvectors $v_n$ are expected to be time independent with the assumption that $C(t)$ is saturated with the lowest $N$ levels in the appropriately chosen values of $t$ and $t_0$. Large time behaviour of the effective mass, given by $m_{\text{eff}}(\tau) = \ln \left(\frac{\lambda_0(\tau + t_0,t_0)}{\lambda_0(\tau + 1 + t_0,t_0)}\right)$, is used to identify plateaus and to choose an optimal fit range, ensuring a reliable extraction of the ground-state energy. Possibility of bound state formation is inferred by comparing the extracted dibaryon mass $E_{D}$ with the corresponding baryon masses, with a necessary condition for binding given by $\Delta E = E_{D} - 2 E_{B} < 0$, where $E_B$ denotes the mass of a single baryon.  

\begingroup
\renewcommand*{\arraystretch}{1.5}
\begin{table*}[htbp]
    \centering
    \begin{tabular}{p{0.1\textwidth}p{0.15\textwidth}p{0.1\textwidth}p{0.15\textwidth}p{0.25\textwidth}}
        \hline
        \hline
        $S_z$& Operator [N] & Spin & Operator [R] & Spin \\
        \hline
        \hline
        +3/2&$\mathcal{O}^N_{1}$ & \{111\}$_S$&$\mathcal{O}^R_{1}$ & \{133\}$_S$\\
        +1/2&$\mathcal{O}^N_{2}$ & \{112\}$_S$&$\mathcal{O}^R_{2}$ & \{233\}$_S$+\{134\}$_S$+\{143\}$_S$ \\
        -1/2&$\mathcal{O}^N_{3}$ & \{122\}$_S$&$\mathcal{O}^R_{3}$ & \{144\}$_S$+\{234\}$_S$+\{243\}$_S$\\
        -3/2&$\mathcal{O}^N_{4}$ & \{222\}$_S$&$\mathcal{O}^R_{4}$ & \{244\}$_S$\\
        \hline
        \hline
    \end{tabular}
    \caption{The spin component $S_z$ of baryon operators ($\mathcal{O}_j^i$) relevant in this work. $\mathcal{O}^i_{j}$ refers to the $j^{th}$ row single baryon operator of $N$ or $R$ embedding of finite volume $H$ irrep. $\{xyz\}_S$ refers to the symmetrized form of the three quark spinor indices $x$, $y$, and $z$, representing the spin structure of the single baryon operator. For more details about operator construction, {\it c.f.} Ref. \cite{Basak:2005ir}.}
    \label{tab:1H_irrep}
\end{table*}
\endgroup

As in other hadronic systems composed of fermions, the Pauli exclusion principle and the antisymmetric color structure dictate that the spin component of single-flavored baryons remains symmetric in the s-wave configuration, suggesting $S = 3/2$. On a finite cubic lattice, spin 3/2 reduces to the $H$ irreducible representation (irrep) \cite{Basak:2005ir}. While a single operator is sufficient to access the ground state, we utilize two operators from independent embeddings of the $H$ irrep to enhance signal plateau and thus ensure a better control over systematics.
These embeddings form the basis of construction for spin components of our $\Omega_{qqq}$ operators, as outlined in Table \ref{tab:1H_irrep}. Based on the Dirac spinor indices in the Dirac-Pauli representation \cite{Basak:2005ir}, the embeddings are classified as nonrelativistic [N] and relativistic [R]. The nonrelativistic embedding [N] utilizes only the components with $\rho = +, s = \pm$ (referring to quark spinor indices 1 and 2), as these survive in the nonrelativistic limit. In contrast, the relativistic embedding [R] includes components with $\rho = -, s = \pm$  (referring to quark spinor indices 3 and 4), which vanish in the nonrelativistic limit due to their leading nontrivial velocity dependence. Here, $\rho$ and $s$ denote the chiral and spin-parity components of the Dirac spinor, respectively \cite{Basak:2005ir}. Note that the total parity of single baryon and dibaryon systems being studied here is positive, and hence is suppressed in the operator descriptions for brevity.

Similar to the case of single-flavor baryons, the Pauli exclusion principle plays a crucial role in deciding the allowed quantum channels in the two-$\Omega$ and two-$\Omega_{ccc}$ systems as well. Generally, two spin-$3/2$ baryons can lead to total spin states $S = 0, 1, 2,$ or $3$. However, the requirement of an antisymmetric wavefunction in an s-wave configuration, combined with a symmetric flavor structure, restricts the system to only two allowed spin channels: $S = 0$ and $S = 2$. This constraint applies to both the $\Omega$-$\Omega$ and $\Omega_{ccc}$-$\Omega_{ccc}$ systems. We will incorporate this symmetry property in constructing operators with the appropriate quantum numbers for the target dibaryon systems. This approach was employed in our previous work \cite{Mathur:2022ovu}, where we studied the maximal-bottom dibaryon system, building upon the framework given in \cite{Basak:2005ir, Buchoff:2012ja}. We briefly discuss the operator construction below. 

As discussed, we consider only $s$-wave interactions in two-baryon systems. The requirement of an antisymmetric wavefunction in this case restricts us to the two allowed spin channels: $S = 0$ and $S = 2$. The dibaryon operator, denoted as $\mathcal{O}_{S_j}^{a,b}$, is designed through direct product of single baryon operators, $\mathcal{O}_{b1}^a$ and $\mathcal{O}_{b2}^b$, followed by projection to the relevant total spins using Clebsch-Gordan (CG) coefficients:
\begin{equation}
    \mathcal{O}_{S_j}^{a,b} = \mathcal{O}_{b1}^a \cdot \text{CG} \cdot \mathcal{O}_{b2}^b, 
    \label{eqn:CG}
\end{equation}
with $S_j$ being the third component of the total spin of the dibaryon system. To access the $S = 0$ and $S = 2$ channels in finite volume, the continuum-based operators are projected onto the corresponding irreducible representations of the octahedral symmetry group. This is achieved through the reduction formula
\begin{equation}
    ^{[S]}\mathcal{O}^{a,b}_{\Lambda,\lambda} = \sum_{S_j}\mathcal{S}_{\Lambda,\lambda}^{S,S_j}\mathcal{O}^{a,b}_{S_j},
    \label{eqn:subduction}
\end{equation}  
where $\mathcal{S}_{\Lambda,\lambda}^{S,S_j}$ represents the reduction coefficients, $\Lambda$ denotes the finite-volume irrep, $\lambda$ refers to its row. In this setup, the $S = 0$ state subduces to the one-dimensional $A_1$ irrep, whereas the five components of the $S = 2$ state are distributed among the two-dimensional $E$ and the three-dimensional $T_2$ irreps \cite{Johnson:1982yq}. Spin structures of the dibaryon operators, constructed following this prescription, are explicitly given in Eq. \eqref{eqn:ops}. The superscripts $a, b$ on $\mathcal{O}$ distinguish between nonrelativistic [N] and relativistic [R] embeddings, as detailed in Table \ref{tab:1H_irrep}. More details about CG coefficients and subduction coefficients are given in Appendix \ref{app:cg}.

\begin{align}
    ^{[0]}\mathcal{O}^{a,b}_{\text{$A_1$,1}} &= \frac{1}{2}\left(\mathcal{O}^a_{1} \mathcal{O}^b_{4} - \mathcal{O}^a_{2} \mathcal{O}^b_{3} + \mathcal{O}^a_{3} \mathcal{O}^b_{2} - \mathcal{O}^a_{4} \mathcal{O}^b_{1}\right),\nonumber\\
    ^{[2]}\mathcal{O}^{a,b}_{\text{$T_2$,1}} &= \frac{1}{\sqrt{2}}\left(\mathcal{O}^a_{1} \mathcal{O}^b_{3} - \mathcal{O}^a_{3} \mathcal{O}^b_{1} \right),\nonumber\\
    ^{[2]}\mathcal{O}^{a,b}_{\text{$T_2$,2}} &= \frac{1}{2}\left(\mathcal{O}^a_{1} \mathcal{O}^b_{2} - \mathcal{O}^a_{2} \mathcal{O}^b_{1} - \mathcal{O}^a_{3} \mathcal{O}^b_{4} + \mathcal{O}^a_{4} \mathcal{O}^b_{3} \right), \label{eqn:ops}\\
    ^{[2]}\mathcal{O}^{a,b}_{\text{$T_2$,3}} &= \frac{1}{\sqrt{2}}\left(\mathcal{O}^a_{2} \mathcal{O}^b_{4} - \mathcal{O}^a_{4} \mathcal{O}^b_{2} \right),\nonumber\\
    ^{[2]}\mathcal{O}^{a,b}_{\text{$E$,1}} &= \frac{1}{2}\left(\mathcal{O}^a_{1} \mathcal{O}^b_{4} + \mathcal{O}^a_{2} \mathcal{O}^b_{3} -\mathcal{O}^a_{3} \mathcal{O}^b_{2} - \mathcal{O}^a_{4} \mathcal{O}^b_{1} \right),\nonumber\\
    ^{[2]}\mathcal{O}^{a,b}_{\text{$E$,2}} &= \frac{1}{2}\left(\mathcal{O}^a_{1} \mathcal{O}^b_{2} - \mathcal{O}^a_{2} \mathcal{O}^b_{1} + \mathcal{O}^b_{3} \mathcal{O}^b_{4} - \mathcal{O}^a_{4} \mathcal{O}^b_{3} \right).\nonumber
\end{align}

\subsection{Wick Contractions}

\begin{figure}[ht]
\includegraphics[width=0.24\textwidth]{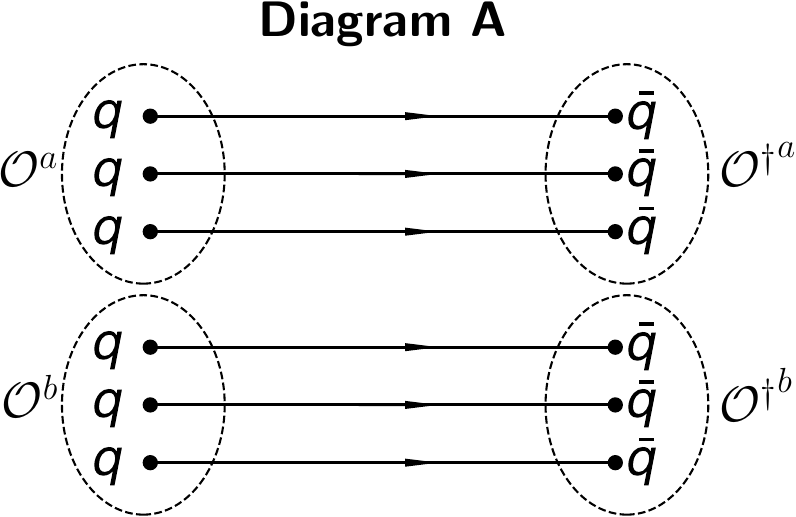}~\vspace*{10pt}
\includegraphics[width=0.24\textwidth]{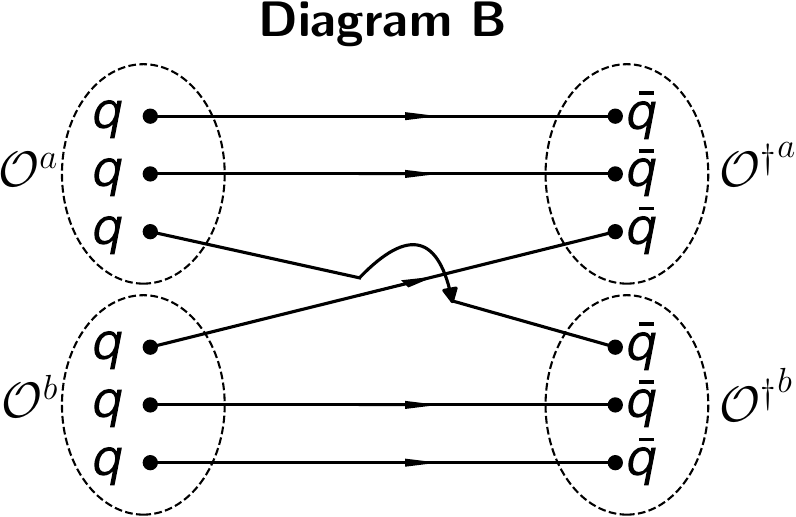}
\includegraphics[width=0.24\textwidth]{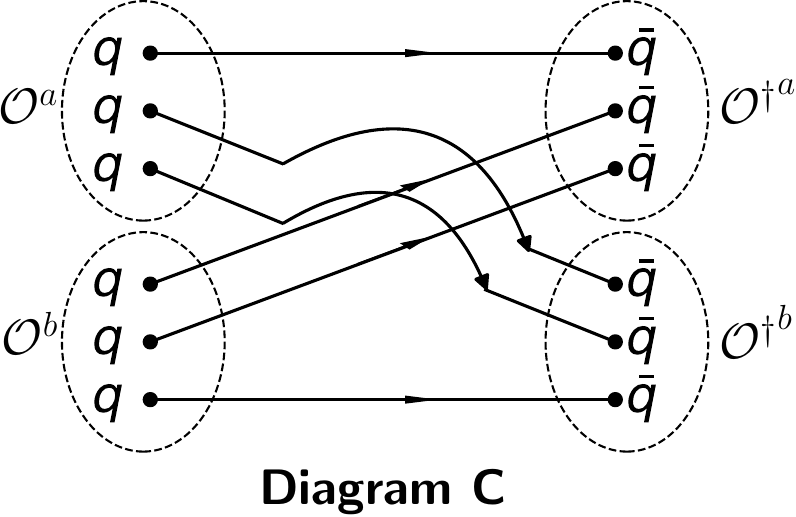}~
\includegraphics[width=0.24\textwidth]{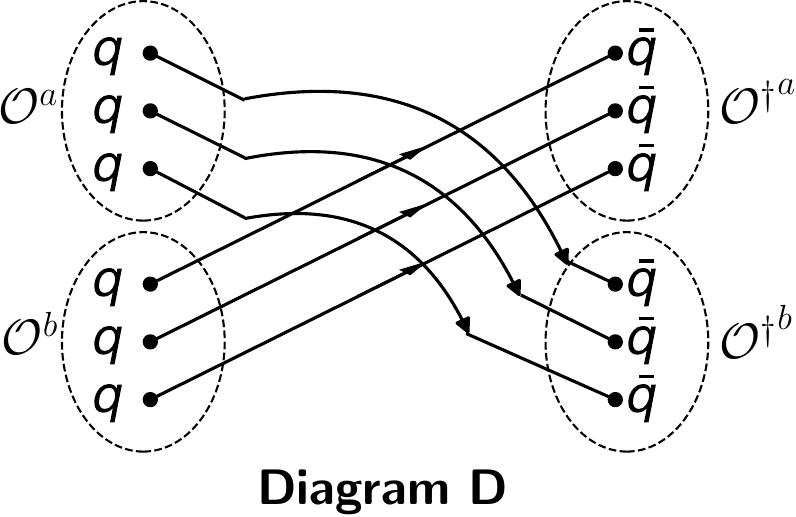}
\caption{\label{fig:cont}Illustration of the four possible distinct Wick contraction diagrams for the $\Omega_{qqq}$-$\Omega_{qqq}$ dibaryon two point correlations. The superscripts $a$ and $b$ on $\mathcal{O}$ denote the single-bayron embeddings used in the operator construction.}
\end{figure}

Given the symmetric flavor structure, the maximum number of distinct Wick contractions in correlation functions between $\Omega_{qqq}-\Omega_{qqq}$ dibaryon operators is four. The possible contractions are shown in Fig. \ref{fig:cont}. Utilizing the two single-baryon embeddings, one can assume the dibaryon interpolator basis of size four. In the spin-0 case, only 3 of these interpolators are linearly independent, forming a correlation matrix of size three. Whereas in the spin-2 case, only 2 linearly independent operators remain. In the spin 2 case, assuming both single-baryons in the [N] embedding at either the source or the sink leads to an effectively vanishing result, stemming from the symmetry constraints imposed by the spin-2 wavefunction, which prevent a purely [N] embedding from coupling effectively to physical states \cite{Buchoff:2012ja}. Diagrams C and D lead to distinct values only when the dibaryon operator is built from a combination of different single-baryon embeddings at the source and at the sink. In all other cases, diagrams C and D become indistinguishable from diagrams B and A, respectively. We utilize these symmetry relations to identify the minimal set of Wick contractions that are required to build the full correlation matrices in our investigation. 

The correlation functions, given in Eq. \eqref{eq:corr}, are constructed for a fixed source timeslice and varying sink time slices. For Spin 0, we have three operators ($^{[0]}\mathcal{O}^{N,N}_{A_1,1}$, $^{[0]}\mathcal{O}^{N,R}_{A_1,1}$, $^{[0]}\mathcal{O}^{R,R}_{A_1,1}$) in the $A_1$ irrep and the corresponding correlation matrix is as follows: 

\begin{widetext}
\begin{equation}
~^{[0]}C_{A_1,1}(t) =
\begin{bmatrix}
\Braket{^{[0]}\mathcal{O}^{N,N}_{A_1,1}\Big|
{^{[0]}\mathcal{O}^{N,N}_{A_1,1}}}~&~
\Braket{^{[0]}\mathcal{O}^{N,N}_{A_1,1}\Big|
{^{[0]}\mathcal{O}^{N,R}_{A_1,1}}}~&~
\Braket{^{[0]}\mathcal{O}^{N,N}_{A_1,1}\Big|
{^{[0]}\mathcal{O}^{R,R}_{A_1,1}}} \\
&&\\
\Braket{^{[0]}\mathcal{O}^{N,R}_{A_1,1}\Big|
{^{[0]}\mathcal{O}^{N,N}_{A_1,1}}} ~&~ 
\Braket{^{[0]}\mathcal{O}^{N,R}_{A_1,1}\Big|
{^{[0]}\mathcal{O}^{N,R}_{A_1,1}}}~&~
\Braket{^{[0]}\mathcal{O}^{N,R}_{A_1,1}\Big|
{^{[0]}\mathcal{O}^{R,R}_{A_1,1}}} \\
&&\\
\Braket{^{[0]}\mathcal{O}^{R,R}_{A_1,1}\Big|
{^{[0]}\mathcal{O}^{N,N}_{A_1,1}}}~&~
\Braket{^{[0]}\mathcal{O}^{R,R}_{A_1,1}\Big|
{^{[0]}\mathcal{O}^{N,R}_{A_1,1}}} ~&~
\Braket{^{[0]}\mathcal{O}^{R,R}_{A_1,1}\Big|
{^{[0]}\mathcal{O}^{R,R}_{A_1,1}}}
\end{bmatrix}.
\label{eq:corr_dibar_s0}
\end{equation}
\end{widetext}

In the case of spin 2, there are five third-components of the spin and hence we can construct five correlation matrices corresponding to the three rows of $T_2$ and two rows of $E$. With two distinct operators for each spin-component ($^{[2]}\mathcal{O}^{N,R}_{\Lambda,\lambda}$, $^{[2]}\mathcal{O}^{R,R}_{\Lambda,\lambda}$), these correlation matrices takes the following form:
\begin{equation}
^{[2]}C_{\Lambda,\lambda}(t) =
\begin{bmatrix} 
\Braket{^{[2]}\mathcal{O}^{N,R}_{\Lambda,\lambda}\Big| {^{[2]}\mathcal{O}^{N,R}_{\Lambda,\lambda}}} ~&~ \Braket{^{[2]}\mathcal{O}^{N,R}_{\Lambda,\lambda}\Big| {^{[2]}\mathcal{O}^{R,R}_{\Lambda,\lambda}}} \\
&\\
\Braket{^{[2]}\mathcal{O}^{R,R}_{\Lambda,\lambda}\Big| {^{[2]}\mathcal{O}^{N,R}_{\Lambda,\lambda}}} ~&~ \Braket{^{[2]}\mathcal{O}^{R,R}_{\Lambda,\lambda}\Big| {^{[2]}\mathcal{O}^{R,R}_{\Lambda,\lambda}}}
\end{bmatrix}.
\end{equation}

In the case of single baryon correlations, the maximal flavor symmetry suggests only a single effective Wick contraction out of six different possible Wick contractions. We build the correlation matrices out of the two operator embeddings list in Table \ref{tab:1H_irrep} as 
\begin{equation}
C_{r}(t) =
\begin{bmatrix} 
\Braket{\mathcal{O}^{N}_{r} \Big| \mathcal{O}^{N}_{r} }&\Braket{\mathcal{O}^{N}_{r}  \Big|\mathcal{O}^{R}_{r}} \\
&\\
\Braket{\mathcal{O}^{R}_{r}  \Big|\mathcal{O}^{N}_{r}} & \Braket{\mathcal{O}^{R}_{r}  \Big|\mathcal{O}^{R}_{r}}
\end{bmatrix},
\label{eq:corr_bar}
\end{equation}
where $r$ refers to the row of the corresponding $H$ irrep representing the single baryon.

\subsection{Energy Extraction and Related Systematics}
\begin{figure}[ht]
\includegraphics[width=0.47\textwidth]{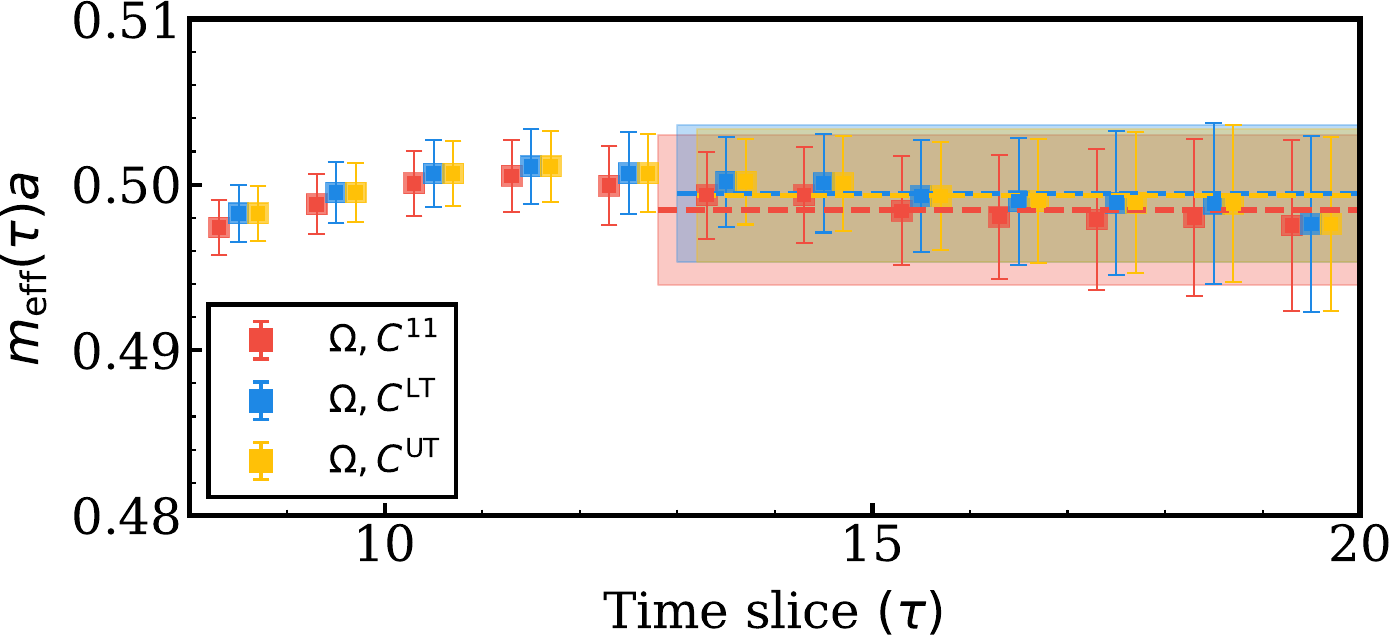}
\includegraphics[width=0.47\textwidth]{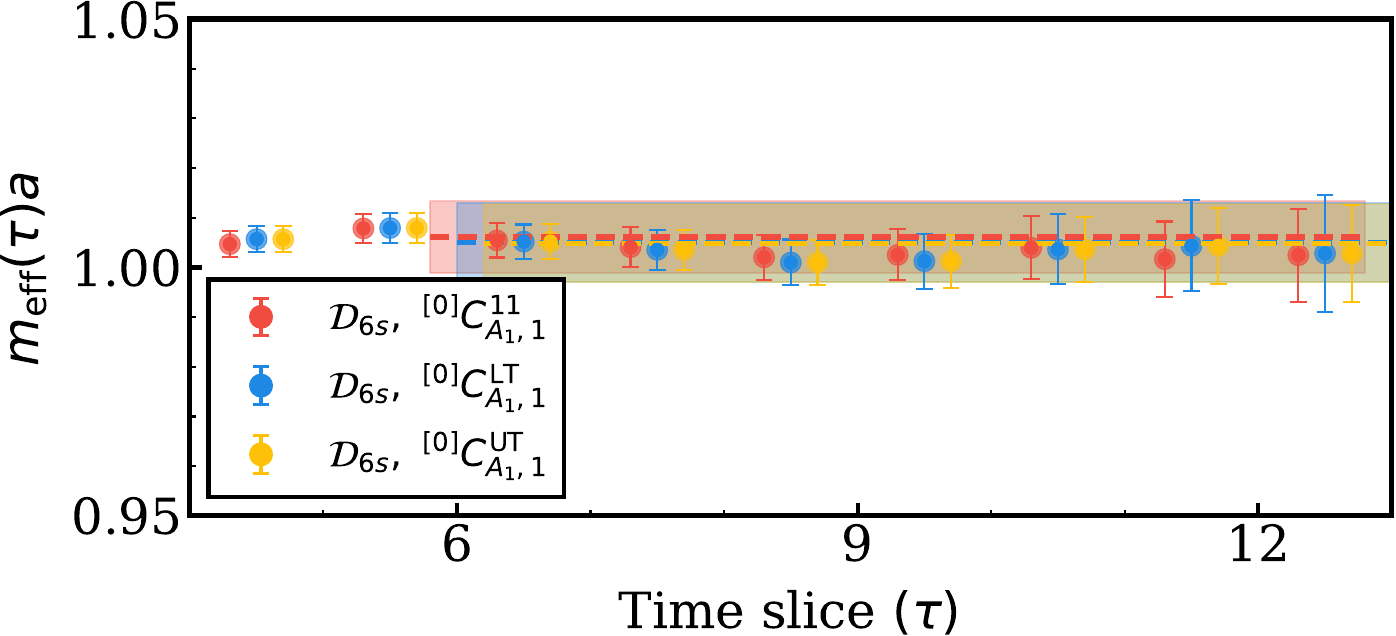}
\caption{\label{fig:corr_dibar_s0} Effective mass plots for the $\Omega$ baryon and its spin-0 dibaryon counterpart on the $S_3$ lattice. Results are shown using only non-relativistic operators correlation function $C_x^{11}$ as well as with the full expanded operator basis, considering both the upper $C_x^{UT}$ and lower $C_x^{LT}$ triangular parts of the correlation matrices as defined in Eq.~\eqref{eq:corr_dibar_s0} and ~\eqref{eq:corr_bar}. }
\end{figure}

In this subsection, we briefly discuss the energy extraction strategy and associated systematics. Although the single non-relativistic operator yields reliable signals for single baryons and spin-zero dibaryons, charm and strange systems may be more strongly influenced by relativistic effects, unlike their comparatively heavier bottom counterparts. With a hope to accommodate these relativistic effects and to enhance the ground state plateau, we incorporate both nonrelativistic and relativistic operators as discussed in the previous subsection, forming an expanded operator basis, and investigate the ground state energies in the baryon and spin-0 dibaryon finite volume spectra. To assess the impact of the expanded operator basis, we present effective mass plots, defined as $m_{\textrm{eff}}(\tau) = \ln \left( \frac{C(\tau)}{C(\tau+1)} \right)$ with $C(\tau)$ being the two-point correlator at source-sink separation $\tau$, on the $S_3$ lattice for both the $\Omega$ baryon and its spin-0 dibaryon counterpart in Figure~\ref{fig:corr_dibar_s0}. The $S_3$ ensemble, being relatively fine, provides a clearer illustration of the effective mass plateauing.

\begin{figure}[ht]
\includegraphics[width=0.47\textwidth]{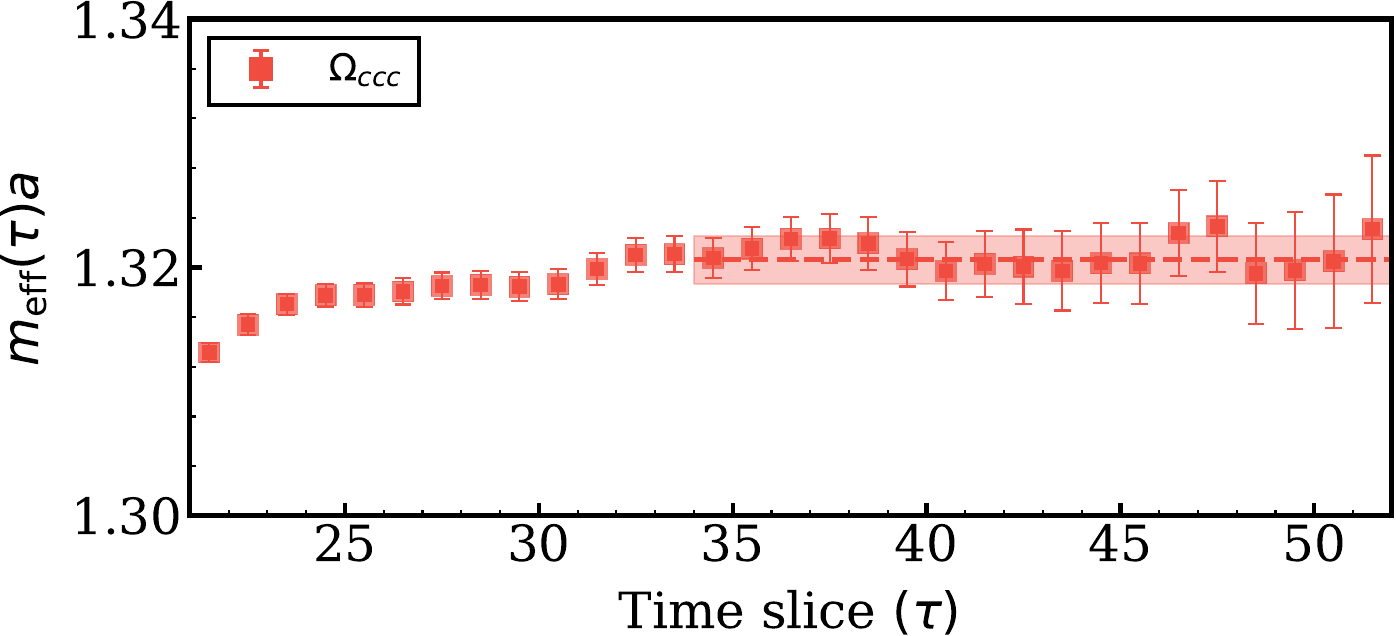}
\includegraphics[width=0.47\textwidth]{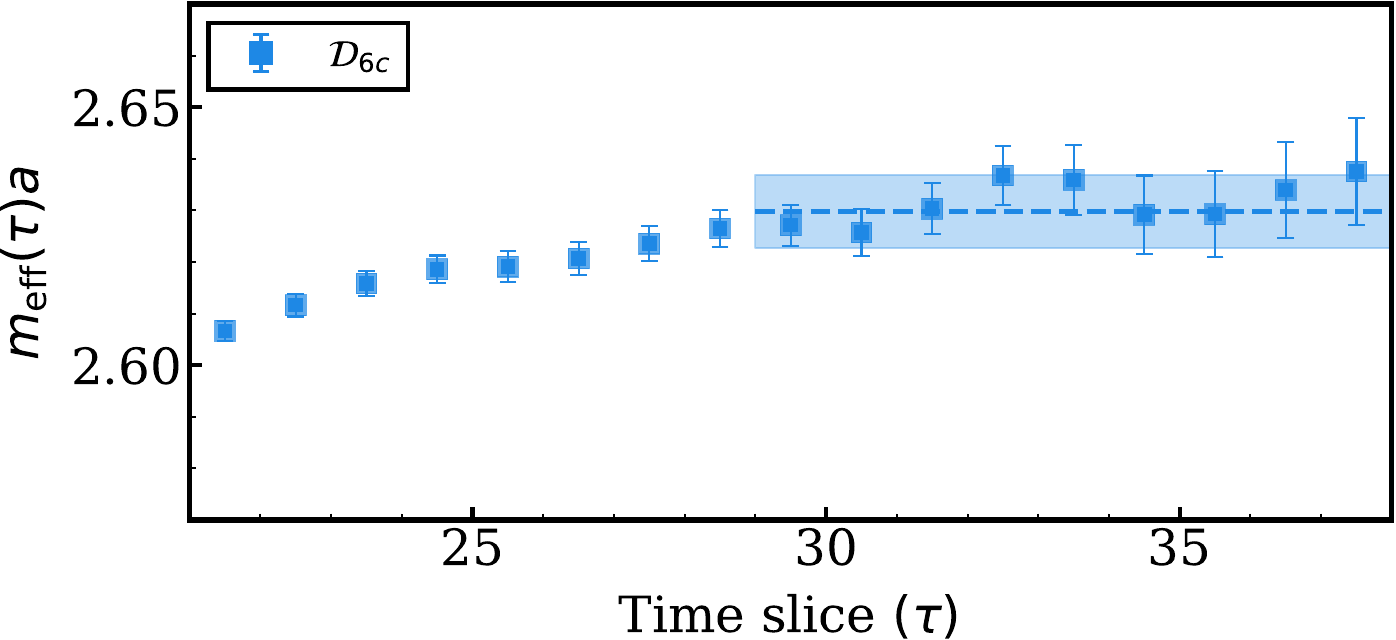}
\includegraphics[width=0.47\textwidth]{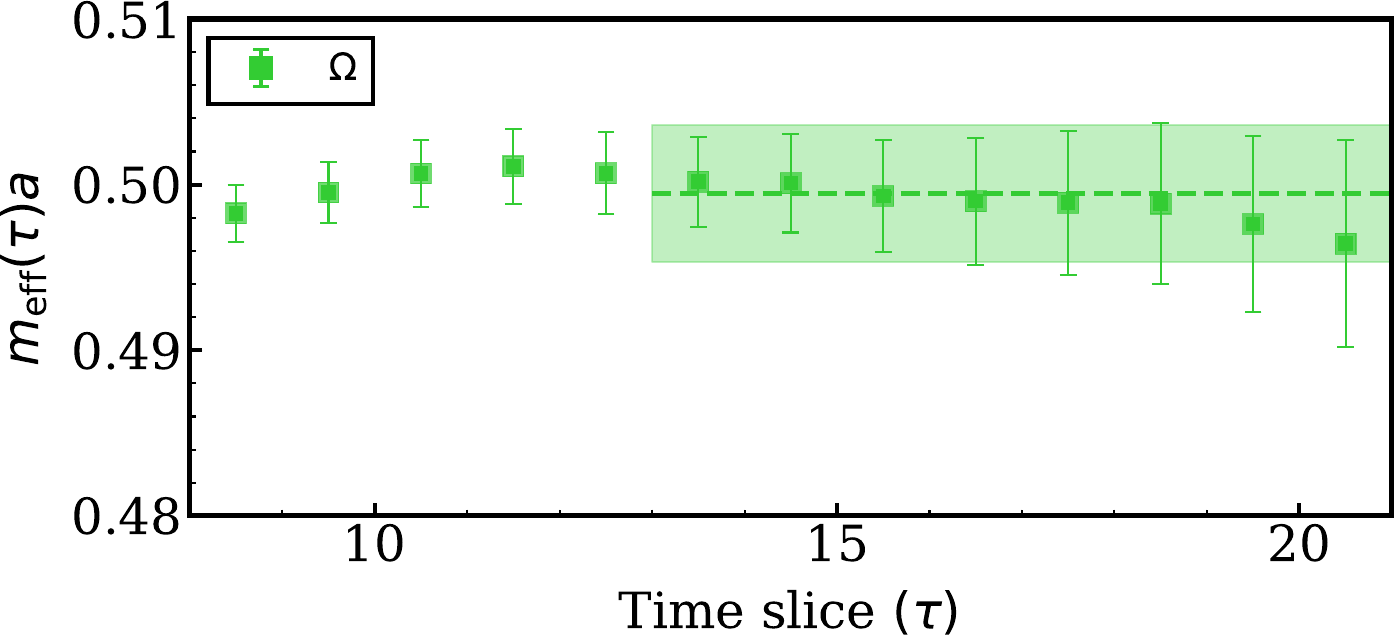}
\includegraphics[width=0.47\textwidth]{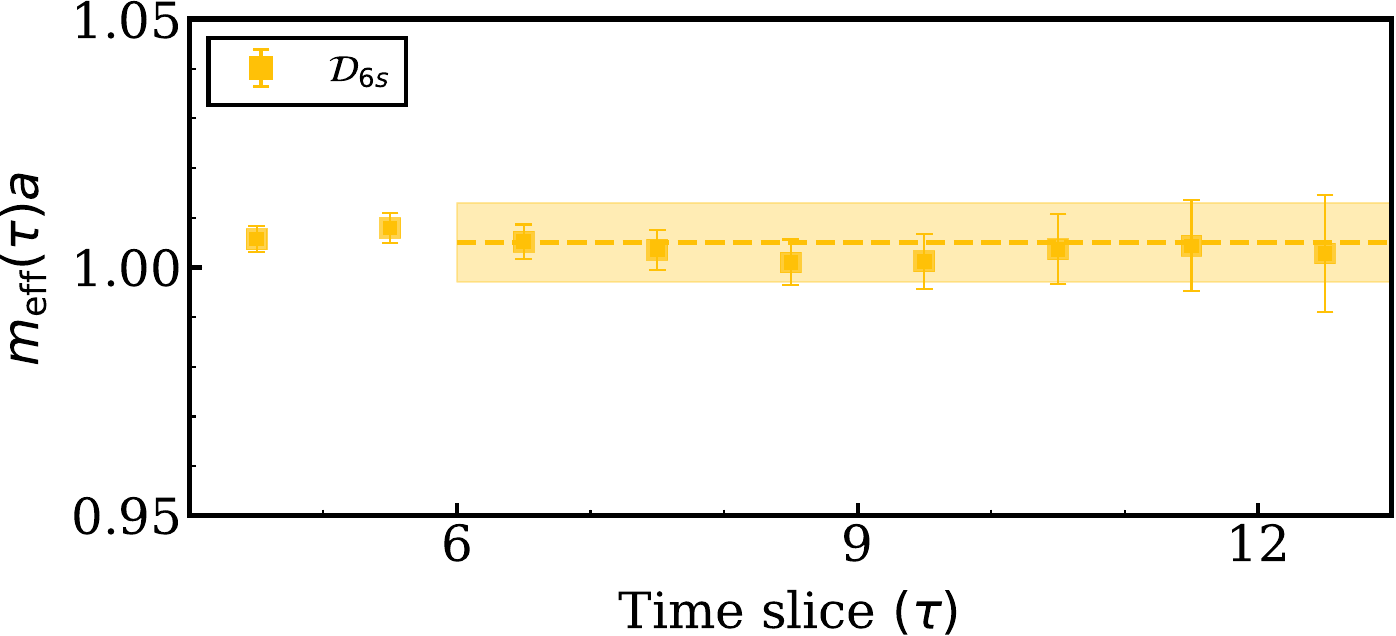}
\caption{\label{fig:l48_effmass} Effective mass plots along with the final energy estimates for the $\Omega_{ccc}$, $\Omega$ baryons and their corresponding Spin-0 dibaryon states $\mathcal{D}_{6c}$ and $\mathcal{D}_{6s}$ on the $S_3$ lattice.}
\end{figure}

We do not observe any significant enhancement in the signal quality, including potential relativistic effects, even for the strange system. This can be observed from the consistency between the mass estimates from single and multioperator bases. To illustrate the consistency in estimates from the upper ($C_x^{\text{UT}}$) and lower ($C_x^{\text{LT}}$) triangular correlations in the asymmetric wall-source to point sink setup (see Eqs. (\ref{eq:corr_bar}) and (\ref{eq:corr_dibar_s0})), we present the effective masses together with correlation functions constructed using only non-relativistic operators ($C_x^{11}$) for the single baryons (top) and spin 0 dibaryons (bottom).  The consistency of mass estimates reflects the remnant Hermiticity properties that we observe in our correlation functions, despite using the asymmetric setup. This was also observed in other systems we have investigated in the past \cite{Mathur:2022ovu,Padmanath:2023rdu,Radhakrishnan:2024ihu}. We also provide the corresponding extracted mass estimates in Table~\ref{tab:corr_dibar_s0}. The effective mass plots along with the final fit choices made for the $\Omega$, $\Omega_{ccc}$, $\mathcal{D}_{6s}$, and $\mathcal{D}_{6c}$ systems on the finer lattice $S_3$ are presented in Fig.~\ref{fig:l48_effmass}. The use of relativistic operators is essential to access the spin-2 dibaryon states, as pure nonrelativistic operators lead to trivial cancellations in this case \cite{Buchoff:2012ja}. 

\begingroup
\renewcommand*{\arraystretch}{1.5}
\begin{table}[htbp]
    \centering
    \begin{tabular}{p{0.1\textwidth}p{0.1\textwidth}p{0.1\textwidth}}
        \hline
        \hline
        CF & $\Omega$ & $\mathcal{D}_{6s}$ (Spin 0) \\
        \hline
        \hline
        $C_x^{11}$ & 0.4985(45) &1.0061(72)\\
        $C_x^{LT}$ & 0.4995(41) &1.0051(80)\\
        $C_x^{UT}$ & 0.4993(40) &1.0048(79)\\
        \hline
        \hline
    \end{tabular}
    \caption{Extracted masses of the $\Omega$ baryon and the $\mathcal{D}_{6s}$ dibaryon in the spin-0 sector, given in lattice units on the $S_3$ lattice. Results are shown using the non-relativistic ([N]) operator basis in the correlation function (CF), and include values obtained from both the upper and lower triangular parts of the respective correlation matrices.}
    \label{tab:corr_dibar_s0}
\end{table}
\endgroup

Taking the example of the charm system, a detailed assessment of the spectral signals from individual operator pairings is presented in Appendix~\ref{app:op_pair}, where instead of using the full correlation matrix, correlation functions constructed from specific pairs of operators are examined. The signal quality is consistently good across all combinations, providing confidence in the robustness of our analysis. All ensembles exhibit similarly reliable signals and are employed in the subsequent investigation of states in the dibaryon systems presented in this work. 

\begin{figure}[ht]
\includegraphics[width=0.47\textwidth]{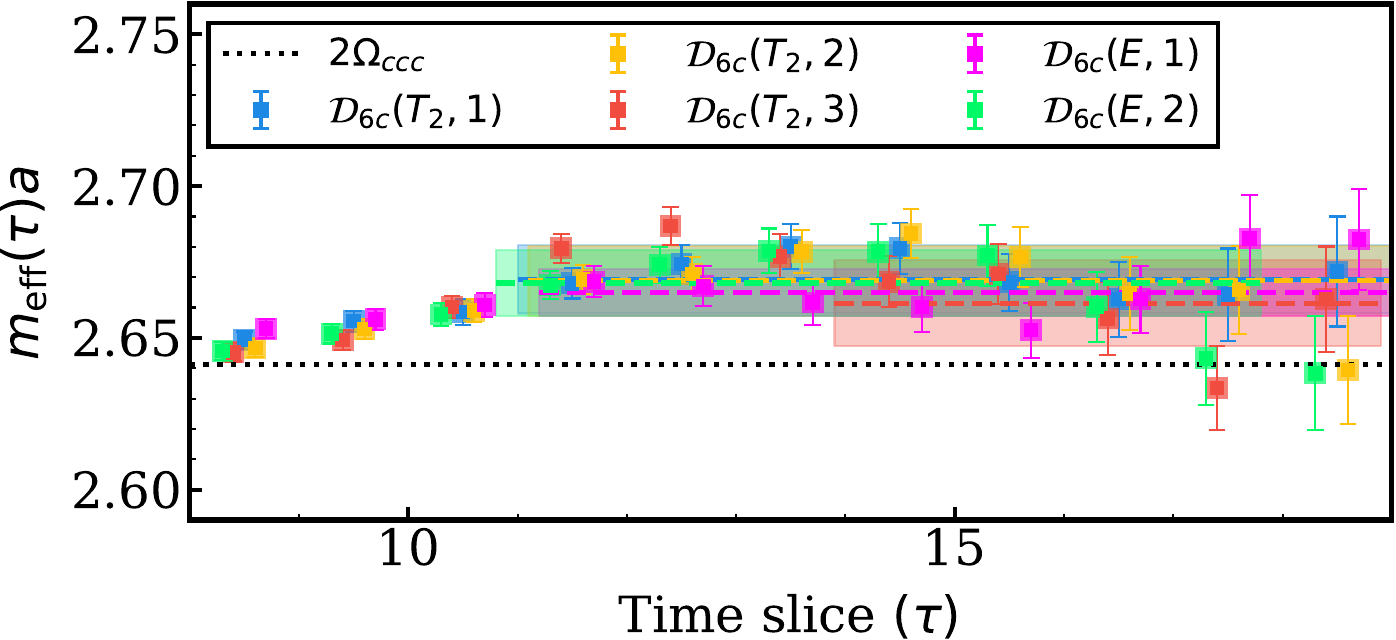}
\includegraphics[width=0.47\textwidth]{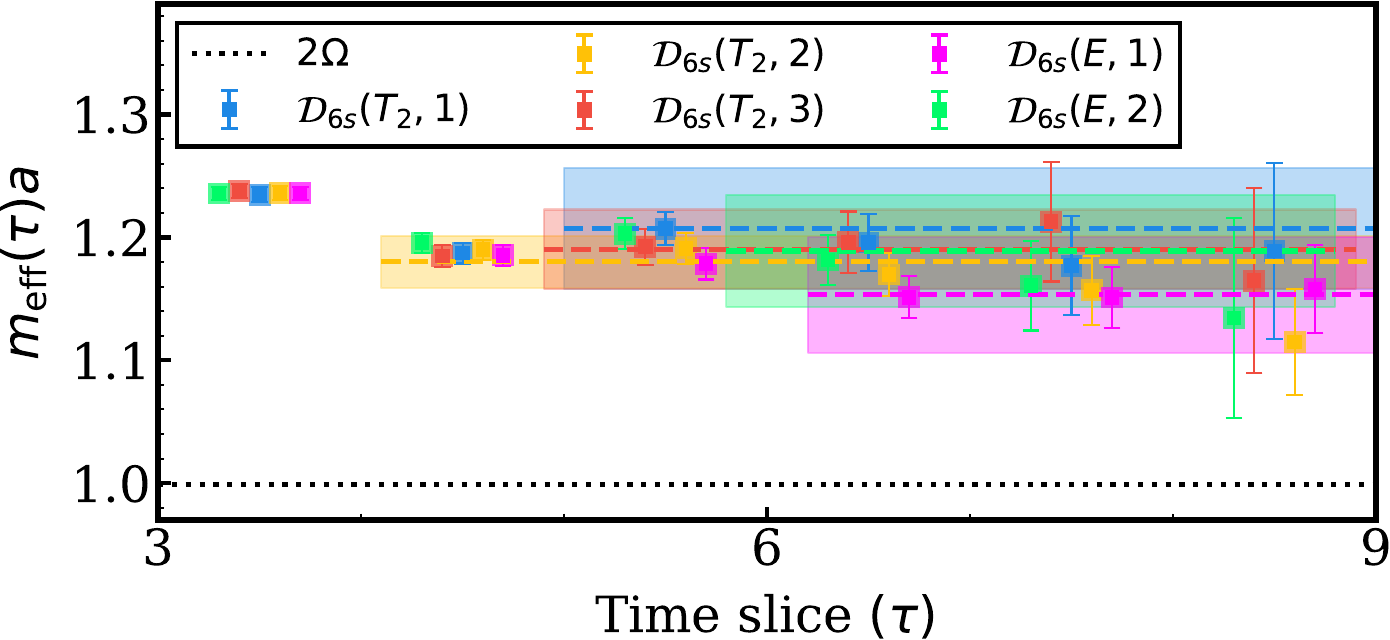}
\caption{\label{fig:spin2_rows} Effective mass plots for the spin-2 dibaryon systems $\mathcal{D}_{6c}$ and $\mathcal{D}_{6s}$ for different rows in the $T_2^+$ and $E^+$ irreps on the $S_3$ lattice, along with the corresponding two-baryon thresholds, read off from Fig.~\ref{fig:l48_effmass}.}
\end{figure}

To assess the effects of reduced finite-volume symmetries on the spin 2 sector, we make a comparative study of corresponding estimates obtained from different rows of the relevant finite-volume irreps. To this end, we examine the consistency in the effective masses extracted from all five rows of the $\mathcal{D}_{6s}$ and $\mathcal{D}_{6c}$ operators in the $T_2$ and $E$ irreps. In Fig.~\ref{fig:spin2_rows}, we present such a comparison made on the $S_3$ lattice, where the quality of the signal allows a reliable comparison across different rows. The fit estimates in the respective cases are presented as bands. A consistent pattern across all five rows can be seen from the Figure, and a similar observation is made throughout all ensembles employed in this study. In this context, we present the relevant two-baryon threshold also in Fig.~\ref{fig:spin2_rows} to gauge the influence of the observed fluctuations with respect to the observed energy shifts. Despite the small fluctuations between different rows of the $T_2$ and $E$ irreps, the mass estimates obtained lie well above the corresponding two-baryon thresholds, as is evident from the figure, and do not qualitatively affect the conclusions we make in the spin 2 sector. The observed consistency points to subdominant reduced symmetry effects on our estimates for the ground states in the spin 2 sector. Accordingly, in the subsequent analysis, we average over these rows in the analysis of the spin 2 dibaryon states. Detailed discussion on the energy shifts observed here is provided later in the next section. 

\begin{figure}[htbp]
\includegraphics[width=0.47\textwidth]{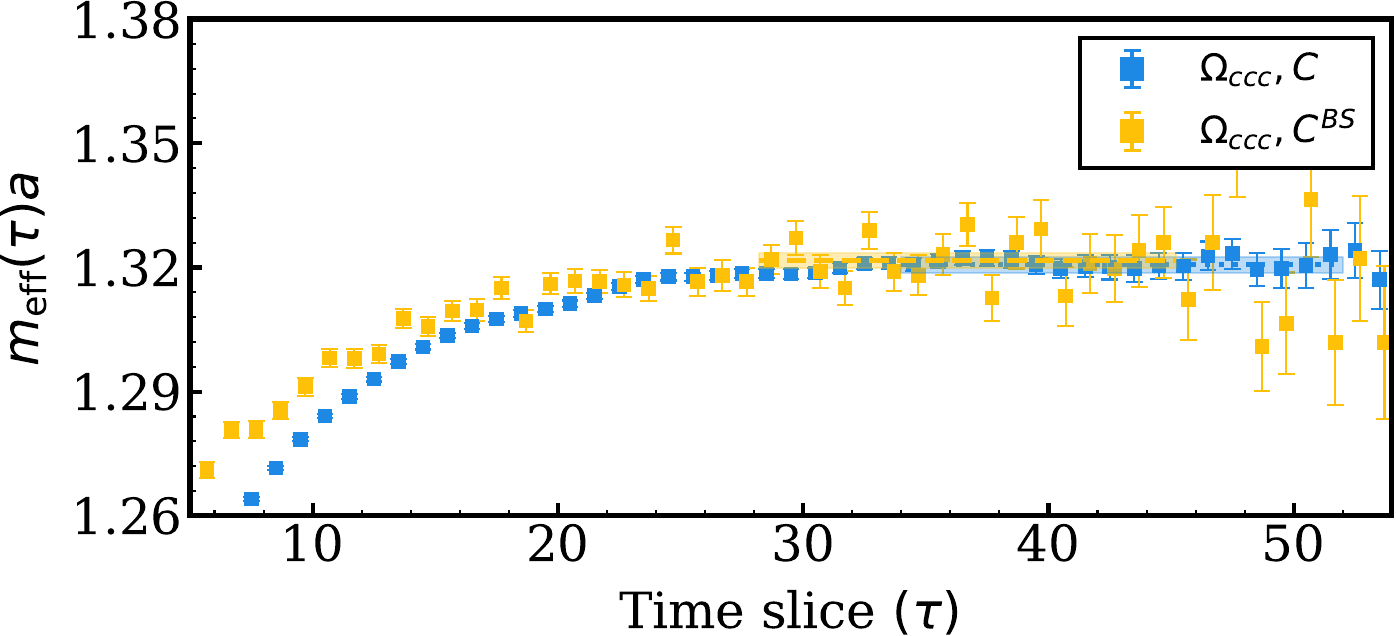}
\includegraphics[width=0.47\textwidth]{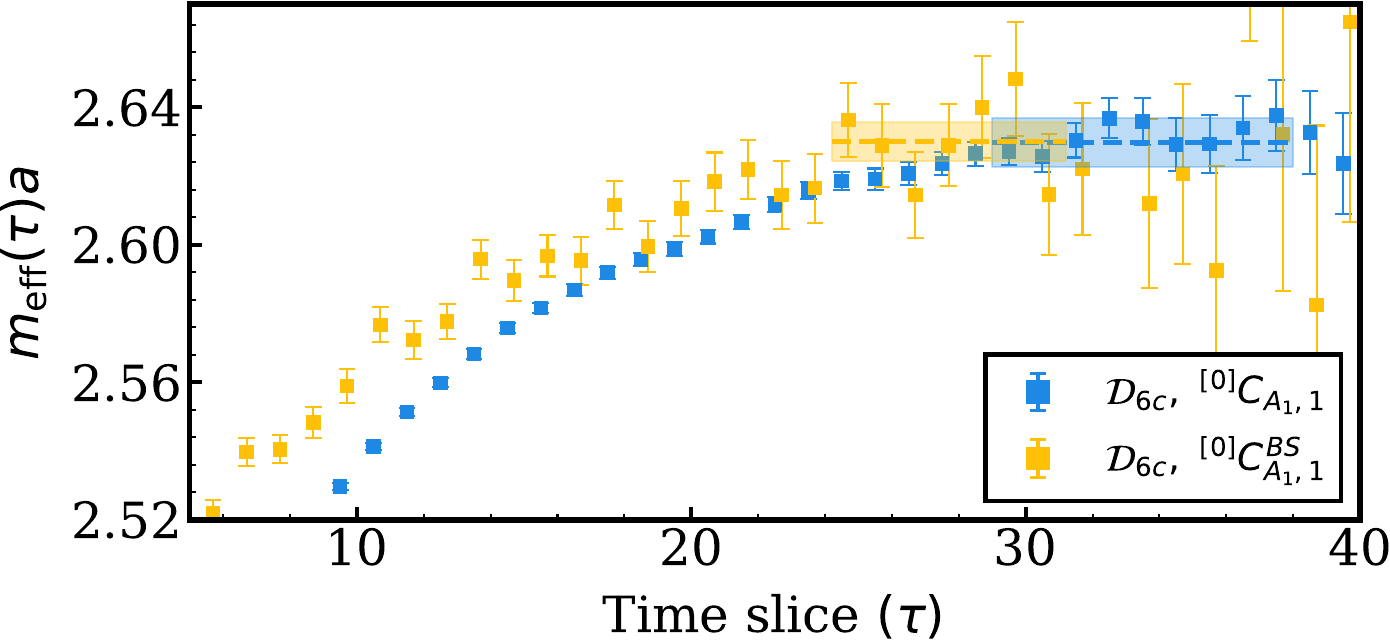}
\caption{\label{fig:sink_structure}Effective mass plots for the $\Omega_{ccc}$ (top) and spin-0 $\mathcal{D}_{6c}$ (bottom) state on the $S_3$ lattice, comparing different sink structures. The box-sink results are with radius 5.}
\end{figure}

Next, we re-examine the identified ground state plateaus by assessing the changes in the approach to the plateau with varying sink smearing procedure. To this end, we compare the ground state estimates determined from the standard wall-to-point setup with those from the box-sink approach. The latter is argued to be an improvised version of the former, wherein the evaluated correlations are expected to have reduced artifacts associated with the asymmetry of wall-to-point correlators~\cite{Hudspith:2020tdf}. This comparison aids in demonstrating the reliability of our extracted ground-state energy estimates that we quote. The extracted energies from the full correlation matrix, using both wall-to-point ($C_x$) and box-sink ($C_x^{BS}$) constructions, remain consistent within statistical uncertainties. For the charmed system on the $S_3$ lattice, this consistency is detailed in Table~\ref{tab:sink_structure} and visualized in Figure~\ref{fig:sink_structure}, confirming the robustness of the extracted ground-state energies across variations in the sink structure. The rise-from-below behavior observed in wall-to-point correlators persists with the box-sink setup, though with reduced magnitude, reflecting a partial reduction in correlator asymmetry.

\begingroup
\renewcommand*{\arraystretch}{1.5}
\begin{table}[htbp]
    \centering
    \begin{tabular}{p{0.1\textwidth}p{0.1\textwidth}p{0.1\textwidth}}
        \hline
        \hline
        CF & $\Omega_{ccc}$ & $\mathcal{D}_{6c}$ (Spin 0) \\
        \hline
        \hline
        $C_x$ & 1.3206(19) & 2.6298(71)\\
        $C_x^{BS}$ & 1.3216(18) & 2.6300(57) \\
        \hline
        \hline
    \end{tabular}
    \caption{Extracted energies for the $\Omega_{ccc}$ and spin 0 $\mathcal{D}_{6c}$ states on the $S_3$ lattice using different sink structures. The consistency across the two setups demonstrates the reliability of the identified ground state plateaus.}
    \label{tab:sink_structure}
\end{table}
\endgroup

We also observe that the use of a multi-operator setup that includes both nonrelativistic and relativistic operators yields effective masses that have reduced effects from the asymmetric setup. A similar behavior is also observed in baryon correlators when relativistic operators are used at both source and sink. This mitigation of the rising-from-below behavior can be observed in the effective mass plots obtained from the matrix elements of the correlation matrices defined in Eqs. (\ref{eq:corr_bar}) and (\ref{eq:corr_dibar_s0}), and illustrated in detail in Appendix \ref{app:op_pair}.  However, the corresponding correlators between relativistic operators alone are noticeably noisier and not suitable to be considered alone for any analysis leading to sensible results. 

%%%%%%%%%%%%%%%%%%%%%%%%%%%%%%%%%%%%%%%%%%%
\section{Results}\label{sec:results}
Following the systematic extraction of finite-volume eigenenergies, we present the key results and observations from our analysis. We present our assessment and inferences from the observed energy splittings between the ground state energies and the associated two-baryon thresholds across all the ensembles. In an elastic system, a negative energy shift with respect to the two-baryon threshold is a clear indication of attractive interaction, whereas a positive energy shift suggests repulsive interaction, assuming no contaminating effects from higher partial waves. We present these energy differences $\Delta E = E_{\text{D}} - 2E_{\text{B}}$ in Tables~\ref{tab:deltaE_spin0} and~\ref{tab:deltaE_spin2} for spin 0 and spin 2 dibaryons, respectively. 

\begingroup
\renewcommand*{\arraystretch}{1.5}
\begin{table}[htbp]
    \centering
    \begin{tabular}{p{0.1\textwidth}p{0.1\textwidth}p{0.1\textwidth}}
        \hline
        \hline
        Ensemble & \multicolumn{2}{c}{$\Delta E$ (Spin 0)} \\
        \cline{2-3}
         & Charm & Strange \\
        \hline
        \hline
        $S_4$ & -54(29) & 22(37) \\
        $S_3$ & -39(28) & 21(38) \\
        $S_2$ & -67(37) & -13(54) \\
        $S_1$ & -70(34) & -86(64) \\
        $L_1$ & -64(29) &  19(64) \\
        \hline
        \hline
    \end{tabular}
    \caption{Summary of extracted energy shifts $\Delta E$ in the spin-0 channel for strange and charmed system.}
    \label{tab:deltaE_spin0}
\end{table}
\endgroup

\begingroup
\renewcommand*{\arraystretch}{1.5}
\begin{table}[htbp]
    \centering
    \begin{tabular}{p{0.1\textwidth}p{0.1\textwidth}p{0.1\textwidth}}
        \hline
        \hline
        Ensemble & \multicolumn{2}{c}{$\Delta E$ (Spin 2)} \\
        \cline{2-3}
         & Charm & Strange \\
        \hline
        \hline
        $S_4$ &  91(24) & 540(41)\\
        $S_3$ & 86(23) & 628(70) \\
        $S_2$ & 24(32) & 318(94)\\
        $S_1$ & 42(35) & 367(75)\\
        $L_1$ & 44(25) & 294(133) \\
        \hline
        \hline
    \end{tabular}
    \caption{Summary of extracted energy shifts $\Delta E$ in the spin-2 channel for strange and charmed system.}
    \label{tab:deltaE_spin2}
\end{table}
\endgroup

A negative energy shift is evident in the spin 0 dibaryons in the charm system with a statistical significance of more than 1$\sigma$ across all the lattices, suggesting an attractive interaction. However, energy shifts in the scalar dibaryons in the strange sector are observed to be consistent with the $\Omega-\Omega$ threshold within 1$\sigma$ errors, suggesting no conclusive evidence for the nature of the underlying interaction. In the spin 2 sector, the energy levels are consistently observed to be positively shifted with respect to the two-baryon thresholds, pointing to the repulsive nature of interactions at both the quark masses studied. 

\begin{figure}[ht]
\includegraphics[width=0.4\textwidth]{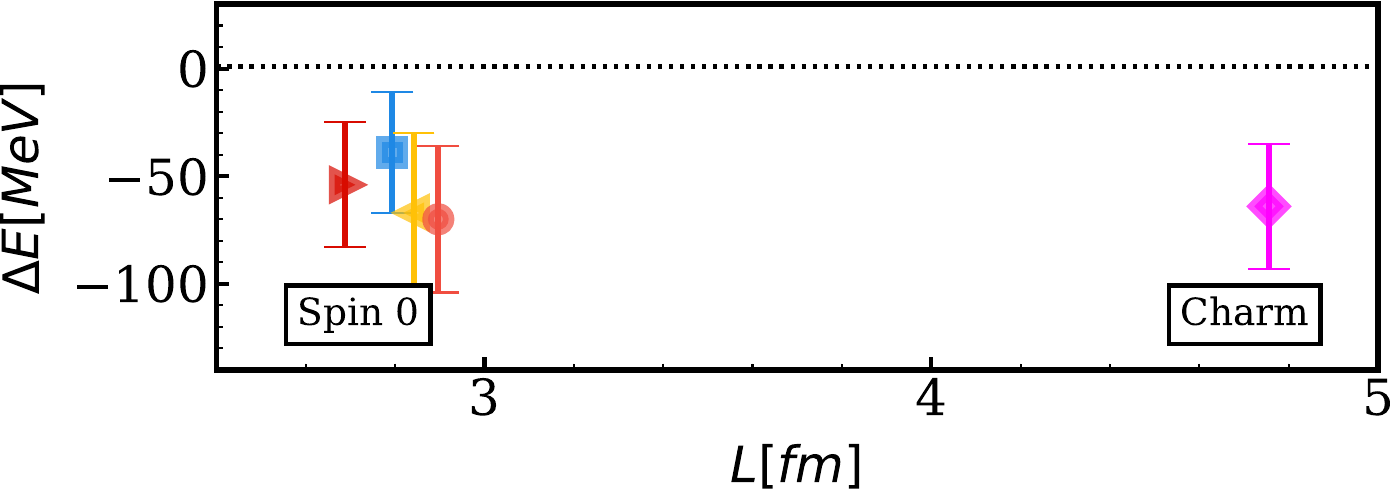}
\includegraphics[width=0.4\textwidth]{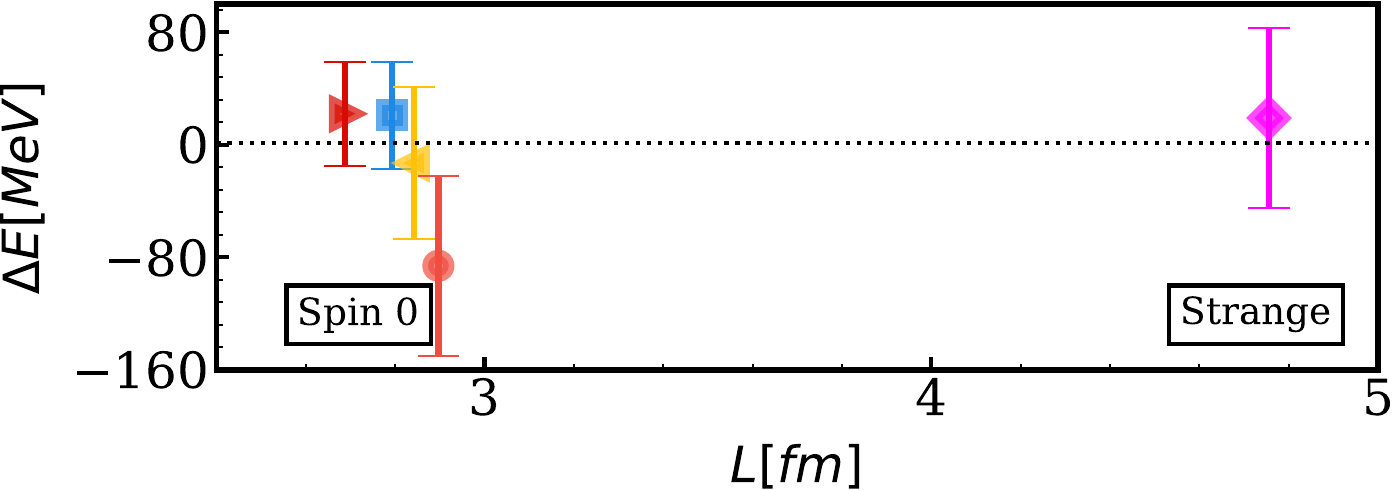}
\includegraphics[width=0.4\textwidth]{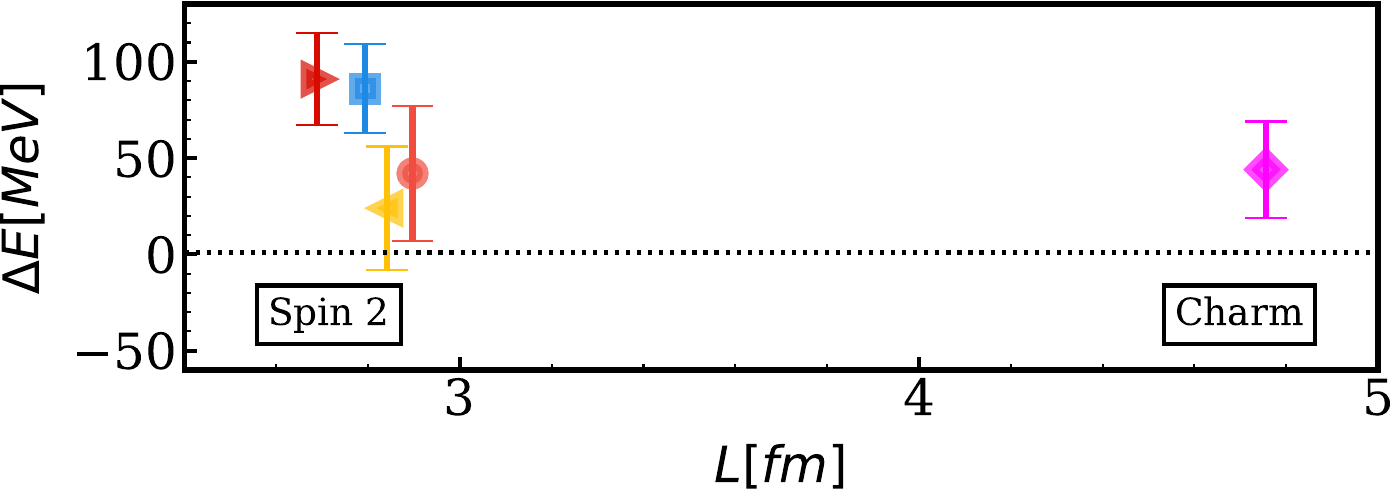}
\includegraphics[width=0.4\textwidth]{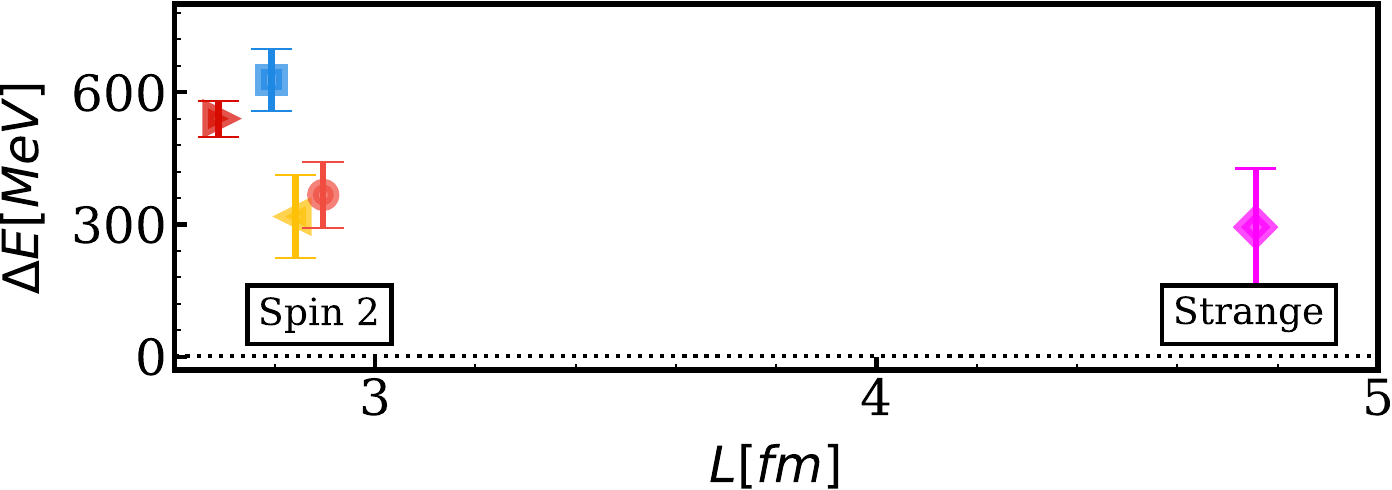}
\caption{\label{fig:ratio} $\Delta E$ plotted against the lattice spatial extent for both charm and strange systems in the spin-0 and spin-2 channels. }
\end{figure}

Now we discuss the volume dependence of the observed energy shifts in the dibaryon systems. We present these energy shifts $
\Delta E$ as a function of spatial lattice extent in Fig.~\ref{fig:ratio}. Observation of consistent negative energy shifts ($\Delta E<0$) across all ensembles in the ground states is indicative of the potentially bound nature of the system. For the spin-0 charmed dibaryon, $\Delta E$ remains negative with little variation across volumes, consistent with a bound state. In contrast, for the spin-0 strange dibaryons, $\Delta E$ shows no clear volume dependence and stays zero within $1\sigma$ uncertainties, making it difficult to draw a definitive conclusion. On the other hand, the energy splittings in spin-2 channels at the strange and charm sectors show positive energy shifts, indicating repulsive interactions and the absence of any bound states. Therefore, we exclude the spin 2 systems from further continuum or amplitude analysis.

\begin{figure}[ht]
\includegraphics[width=0.45\textwidth]{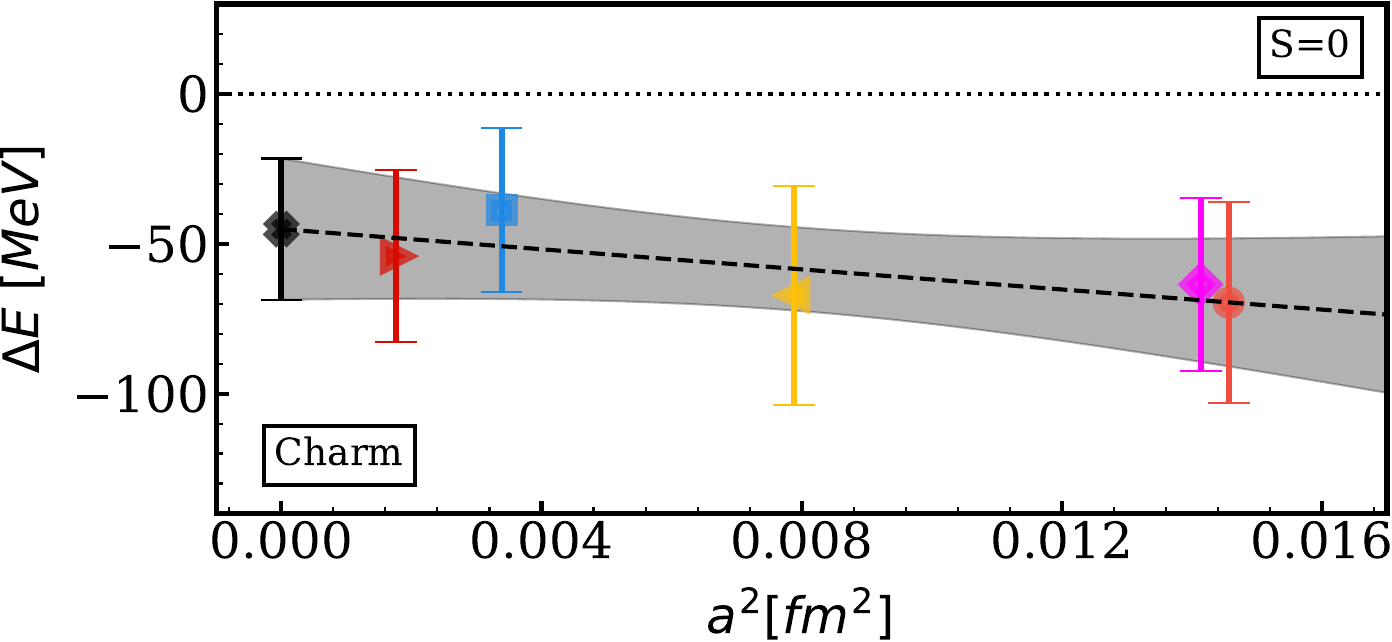}
\includegraphics[width=0.45\textwidth]{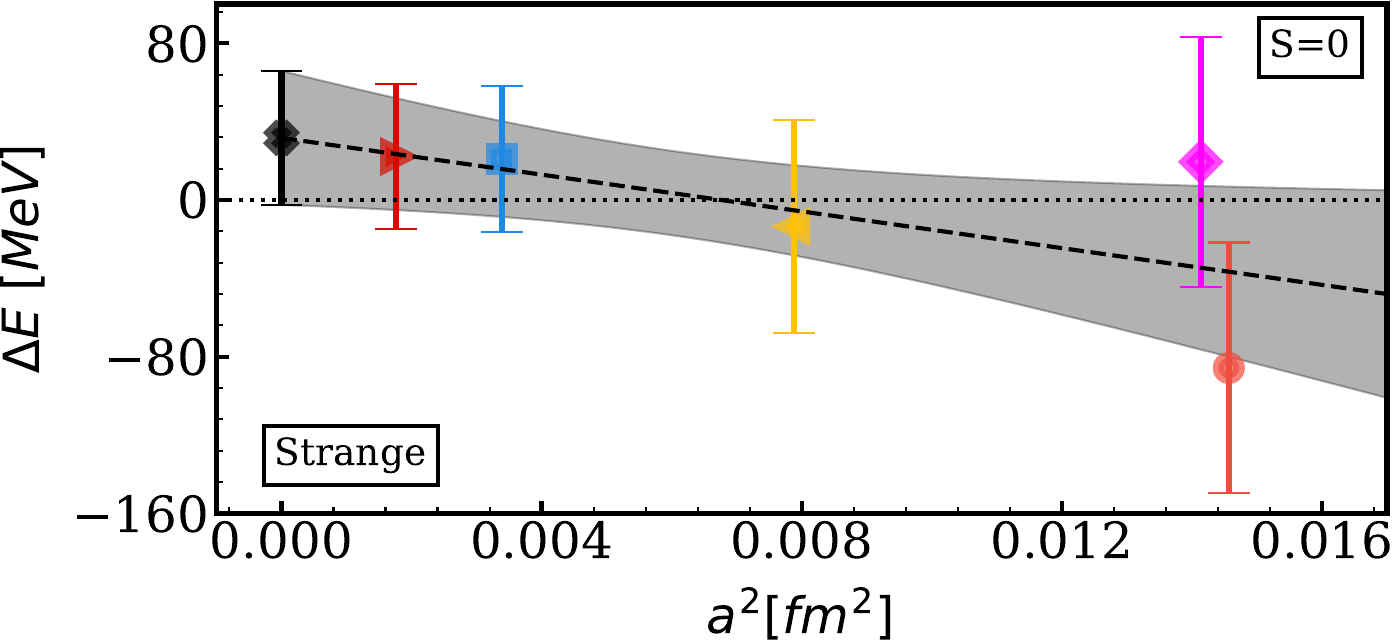}
\caption{\label{fig:fit}     Continuum extrapolation of the energy shifts $\Delta E$ for spin-0 dibaryon systems composed of charm (above) and strange (below) quarks. The energy shifts, expressed in MeV, are plotted as a function of the lattice spacing squared ($a^2$) and fitted using the form $f(a) = c_0 + c_1 a^2$. The extrapolated values at $a \to 0$ represented by the black data point provide a continuum estimate of the binding behavior in each case.}
\end{figure}

Being composed of heavy quarks, the charmed scalar dibaryons could be subject to strong discretization effects. To address these effects, these energy shifts are fit to the functional form $f(a) = c_0 + c_1 a^2$, with errors determined following a bootstrap procedure. The continuum extrapolation enables a more definitive assessment of potential binding in the $\mathcal{D}_{6q}$ system at the charm and strange points based on the observed energy shifts at finite lattice spacing. The resulting extrapolations are presented in Fig.~\ref{fig:fit}. The continuum value for binding energy in the charmed system is found to be $-45(24)$ MeV. In contrast, the strange system yields a continuum value of $32(34)$ MeV, suggesting a ground state energy consistent with the energy of the two-baryon threshold. We also perform a more appropriate finite-volume treatment {\it \'a la} L\"uscher \cite{Luscher:1990ux,Briceno:2014oea} for $\Omega_{qqq}\Omega_{qqq}$ scattering using these extracted ground state energies and with a zero range approximation for the amplitude supplemented with a leading quadratic lattice spacing dependence presented above. The resultant binding energies in the charm and strange sectors are consistent with the continuum extrapolated numbers we have quoted above. Assessing these results based on the observed quark mass dependence, they are qualitatively consistent with our earlier lattice study of bottom-flavored dibaryon \cite{Mathur:2022ovu}. For the strange system, our near-threshold $\Delta E$ aligns with previous findings of weak attraction or repulsion in the $\Omega$-$\Omega$ system \cite{Buchoff:2012ja,HALQCD:2015qmg,Gongyo:2017fjb}. In the charm sector, our observation supports the possibility of a bound $\Omega_{ccc}$-$\Omega_{ccc}$ state, qualitatively similar to the conclusions in Ref.~\cite{Lyu:2021qsh} in the pure QCD scenario, albeit with a deeper binding (with a large uncertainty) in our case. Assuming a molecular picture for the $\Omega_{ccc}$-$\Omega_{ccc}$ system with a separation of 1.1fm \cite{Liu:2021pdu}, electrostatic effects will reduce the observed binding further by about 10 MeV, and even more, if it were a more compact system. The $\Omega$-$\Omega$ would be less influenced by such electrostatic effects owing to the smaller total electric charge and a relatively large size of the system, with the expected shift being of the order of 1 MeV, as also observed in Ref.~\cite{Gongyo:2017fjb}. In the spin-2 sector, the binding energies obtained for the charmed and strange dibaryon systems remain more than $3\sigma$ above threshold in the continuum limit, indicating repulsive interactions in the tensor channels.

\begin{figure}[ht]
\includegraphics[width=0.47\textwidth]{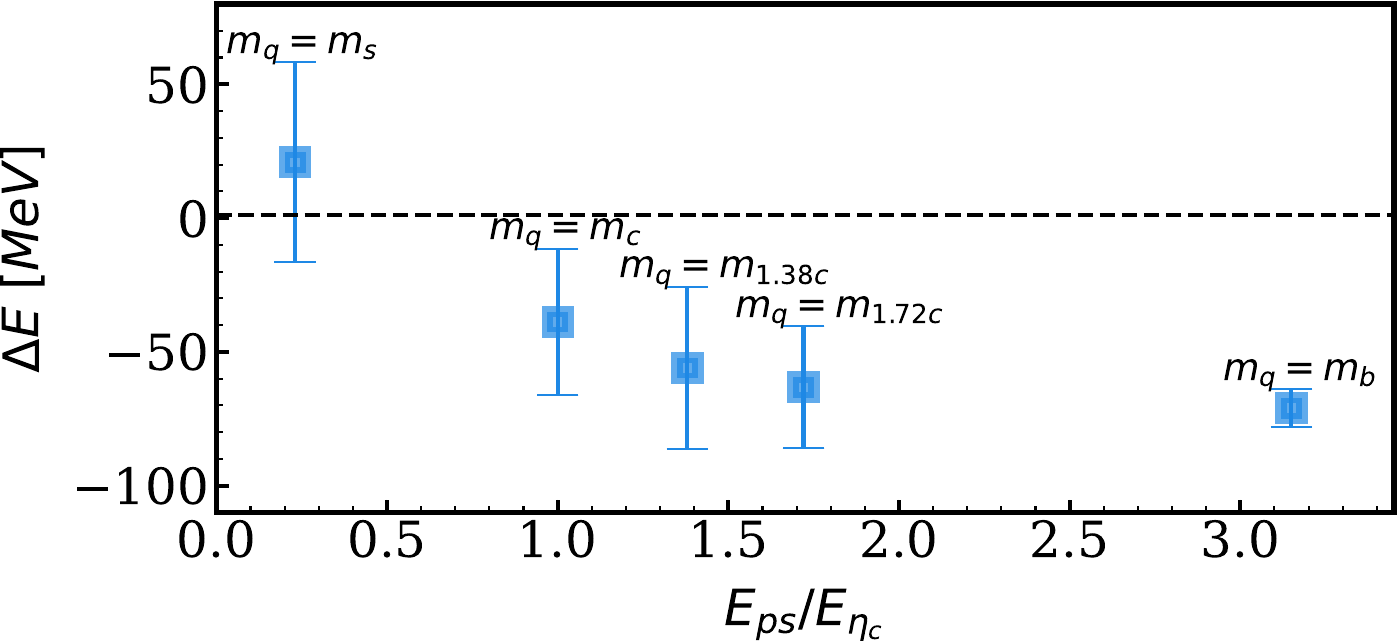}
\caption{\label{fig:charm_mass} Energy shift $\Delta E$ in the spin-0 channel for the $\Omega_{qqq}$-$\Omega_{qqq}$ system as a function of $E_{ps}/E_{\eta_c}$ on the $S_3$ lattice. The trend shows increasing attraction with heavier valence quark mass. The result for $m_q=m_b$ is taken from Ref. \cite{Mathur:2022ovu}.}
\end{figure}

To further quantify the observed quark mass dependence of the spin-0 dibaryon ground states, we analyze the energy splitting between the dibaryon and the corresponding two-baryon threshold on the finest lattice, $S_3$, by varying the valence quark mass. Specifically, we investigate four different quark masses: $m_q = m_s$ (physical), $m_q = m_c$ (physical), $m_q = m_{1.38c}$, and $m_q = m_{1.72c}$, and compare our numbers with those obtained at the bottom quark mass $m_q=m_b$ (physical) \cite{Mathur:2022ovu}. As the quark mass increases, we observe a clear trend of stronger binding of the spin-0 dibaryon state with respect to the non-interacting threshold, indicating an increasingly attractive nature of interactions at the heavier quark masses. This observation is consistent with our previous investigation of bottom-flavored dibaryons, where stronger binding was observed \cite{Mathur:2022ovu}. The corresponding energy shifts for different quark masses are summarized in Table~\ref{tab:charm_mass} and illustrated in Figure~\ref{fig:charm_mass}.

\begingroup
\renewcommand*{\arraystretch}{1.5}
\begin{table}[ht]
	\centering
	\begin{tabular}{cc} \hline \hline 
	~~$m_q$~~ &     ~~$\Delta E$ (MeV)~~\\ \hline \hline  
    $m_s$ &   21(38)  \\
    $m_c$ &   -39(27)  \\
    $m_{1.38c}$ &   -56(30) \\
    $m_{1.72c}$ &  -63(23) \\ 
    $m_b$~\cite{Mathur:2022ovu} & -71(7) \\ \hline \hline
	\end{tabular}
	\caption{\label{tab:charm_mass}{Energy shift $\Delta E = M_{\text{D}} - 2M_{\text{B}}$ in the spin-0 channel for the $\Omega_{qqq}$-$\Omega_{qqq}$ system on the $S_3$ lattice, shown for four different quark masses: $m_q = m_s$, $m_q = m_c$, $m_q = m_{1.38c}$, and $m_q = m_{1.72c}$. The corresponding estimate at the bottom quark mass is taken from Ref. \cite{Mathur:2022ovu}. The increasingly negative values of $\Delta M$ with heavier quark masses indicate stronger binding.}}
	\end{table} 
\endgroup

%%%%%%%%%%%%%%%%%%%%%%%%%%%%%%%%%%%%%%%%%%%
\section{Summary and Discussion}\label{sec:conc}
\begin{figure*}[ht]
\resizebox{\linewidth}{!}{\input{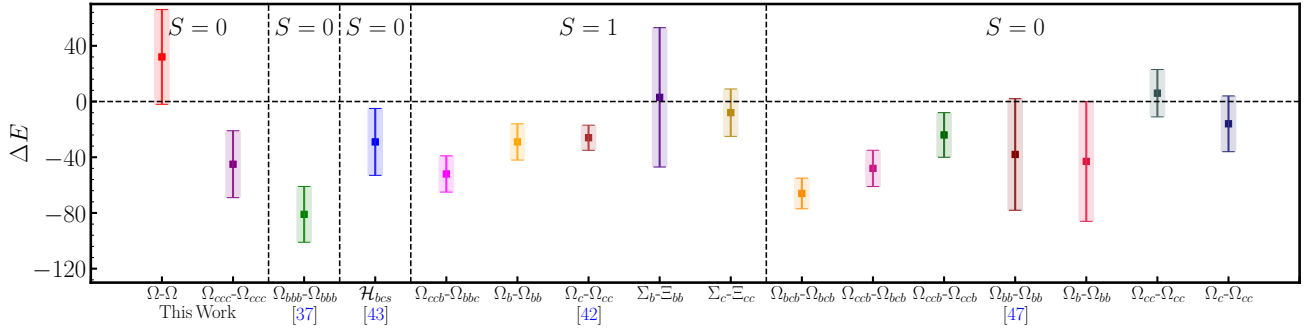}}
\caption{\label{fig:final} Comparison of our results for the $\Omega$–$\Omega$ and $\Omega_{ccc}$–$\Omega_{ccc}$ systems with our previous studies on heavy dibaryon candidates \cite{Mathur:2022ovu, Junnarkar:2019equ, Junnarkar:2022yak, Junnarkar:2024kwd}. While the $\Omega$–$\Omega$ system, composed of six strange quarks, is not typically categorized as heavy, it is included here for completeness and direct comparison. The plotted values represent energy shifts of the dibaryon states relative to their respective non-interacting two-baryon thresholds in different spin channels. The labeling of the systems in the figure is consistent with the notation used in the respective source references.}
\end{figure*}

In this work, we present a lattice QCD investigation of dibaryon systems with maximal charm and strangeness in spin-0 and spin-2 channels. The study is performed using five lattice QCD ensembles with $N_f = 2+1+1$ dynamical flavors generated using the Highly Improved Staggered Quark (HISQ) action. These ensembles span four distinct lattice spacings and include two different spatial volumes, enabling a systematic assessment of discretization and finite-volume effects. In the valence sector, the strange and charm quarks are treated using the overlap action.

To extract ground state energies reliably, we employ a carefully designed operator basis in the evaluation of correlation matrices for single baryons and dibaryons, and implement a variational program following the solutions of a Generalized Eigenvalue Problem (GEVP). The robustness of our ground state energy extraction is demonstrated through consistent effective mass plateaus across different operator combinations. Additionally, we validate our ground state mass plateau identification by comparing the standard wall-to-point correlators with an alternative box-sink setup, both yielding consistent estimates within uncertainties. We summarize the main findings of our analysis below: 

\begin{itemize}
    \item The charmed scalar dibaryon system consistently exhibits a negative energy shift across all ensembles, indicating attractive interactions. A continuum extrapolation of the observed energy shifts yields a binding energy of $\Delta E = -45(24)$ MeV, supporting the presence of a bound state. The inclusion of electrostatic effects reduces the strength of this binding by 10 MeV. Furthermore, by varying the valence charm quark mass on a fine $S_3$ lattice, we observe that the attraction strengthens with increasing quark mass, reinforcing the trend seen in previous heavy-quark dibaryon studies.
    \item In the spin-0 channel, the strange system exhibits energy shifts that are consistent with zero within 1$\sigma$ uncertainties. The continuum extrapolation yields a binding energy of $\Delta E = 32(34)$ MeV, indicating only a weak interaction. In this case, the inclusion of electrostatic effects leads to a mild reduction in the binding energy, by about 1 MeV, which does not significantly alter the qualitative picture.
    \item In the spin-2 channel, both the strange and charm dibaryon systems exhibit consistent positive energy shifts, suggesting repulsive behavior across most ensembles, with no indication of binding. This trend is consistently observed across all operator rows of the $T_2$ and $E$ irreps and across all lattice spacings and spatial volumes. The repulsion is more pronounced in the strange sector compared to the charm system.
\end{itemize}

In our previous works, we explored a range of dibaryon candidates, and it is useful to compare the current findings with those earlier results. For this purpose, we present a summary in Table~\ref{tab:final} and Figure~\ref{fig:final}, showcasing various dibaryon systems across different spin channels. These candidates range from triply heavy to fully heavy dibaryons, as studied in Refs.~\cite{Mathur:2022ovu, Junnarkar:2019equ, Junnarkar:2022yak, Junnarkar:2024kwd}. Although the strange dibaryon examined in this work does not fall under the heavy dibaryon category, it is included here for completeness and comparative context.

\begingroup
\renewcommand*{\arraystretch}{1.5}
\begin{table}[ht]
	\centering
	\begin{tabular}{cccc} \hline \hline 
	~~Dibaryon~~ & ~~Ref.~~     & ~~Spin~~ &~~$\Delta E$ (MeV)~~\\ \hline \hline    
    $\Omega$-$\Omega$ & This Work  &  $S=0$&32(34) \\
	$\Omega_{ccc}$-$\Omega_{ccc}$ &   &  & -45(24) \\  \hline 
	$\Omega_{bbb}$-$\Omega_{bbb}$ & \cite{Mathur:2022ovu} & $S=0$ & -81(16)(14)  \\ \hline 
	$\mathcal{H}_{bcs}$ & \cite{Junnarkar:2022yak}& $S=0$ &  -29(24)  \\ \hline 
    $\Omega_{ccb}$-$\Omega_{bbc}$ &  &  & -52(13)  \\
    $\Omega_{b}$-$\Omega_{bb}$ &  &  &  -29(13) \\
    $\Omega_{c}$-$\Omega_{cc}$ & \cite{Junnarkar:2019equ} & $S=1$ &  -26(9) \\
    $\Sigma_{b}$-$\Xi_{bb}$ &  &  & 3(50)  \\
    $\Sigma_{c}$-$\Xi_{cc}$ &  &  &  -8(17) \\ \hline
    $\Omega_{bcb}$-$\Omega_{bcb}$ &  &  &  -66(11) \\
    $\Omega_{ccb}$-$\Omega_{bcb}$ &  &  &  -48(13) \\
    $\Omega_{ccb}$-$\Omega_{ccb}$ &  &  &  -24(16) \\
    $\Omega_{bb}$-$\Omega_{bb}$ & \cite{Junnarkar:2024kwd} & $S=0$ &  -38(40) \\
    $\Omega_{b}$-$\Omega_{bb}$ &  &  &  -43(43) \\
    $\Omega_{cc}$-$\Omega_{cc}$ &  &  &  6(17) \\
    $\Omega_{c}$-$\Omega_{cc}$ &  &  &  -16(20) \\ \hline \hline
	\end{tabular}
	\caption{\label{tab:final}{Summary of our results for various dibaryon systems, including $\Omega$–$\Omega$ and $\Omega_{ccc}$–$\Omega_{ccc}$ studied in this work, along with earlier heavy dibaryon candidates from Refs.~\cite{Mathur:2022ovu, Junnarkar:2019equ, Junnarkar:2022yak, Junnarkar:2024kwd}. The energy shift for each dibaryon system is listed and also illustrated in Fig.~\ref{fig:final}. The notation used for the listed systems is adapted from the respective references from which the corresponding results are drawn.
}}
	\end{table} 
\endgroup

Our results highlight the potential existence of bound states in the charmed scalar dibaryon sector, while strange dibaryons appear to be near the scattering threshold. To establish these conclusions more firmly, future high statistics studies involving detailed finite-volume analyses will be essential to clarify the pole patterns in the respective amplitudes and their residues in these channels. Additionally, further theoretical investigations employing effective field theories and phenomenological models across a range of quark masses, from light to heavy, would provide valuable insights into how quark mass variations influence dibaryon formation and binding dynamics.

%%%%%%%%%%%%%%%%%%%%%%%%
\acknowledgments
The authors would like to thank Debsubhra Chakraborty, Tanishk Shrimal, and Bhabani Sankar Tripathy for valuable discussions. This work is supported by the Department of Atomic Energy, Government of India, under Project Identification Number RTI 4002. We are thankful to the MILC collaboration and, in particular, to S. Gottlieb for providing us with the HISQ lattice ensembles. Computations were carried out on the Cray-XC30 of ILGTI, TIFR (which has recently been closed), and the computing clusters at DTP, TIFR Mumbai, and IMSc Chennai. NSD would also like to thank Anusandhan National Research Foundation (ANRF), India, for the International Travel Support (ITS/2025/000390).  MP gratefully acknowledges support from the Department of Science and Technology, India, through the ANRF (previously known as SERB) Start-up Research Grant No. SRG/2023/001235. We thank the authors of Ref.~\cite{Morningstar:2017spu}
for making the \textit{TwoHadronsInBox} package public.
%%%%%%%%%%%%%%%%%%%%%%%%%%%%%%%%%%%%%%%%%%%
\appendix
\renewcommand{\thefigure}{\Alph{figure}}
\setcounter{figure}{0}
\section{Detailed Spectral Results by Operator Pairing}\label{app:op_pair}
\begin{figure*}[ht]
\includegraphics[width=0.32\textwidth]{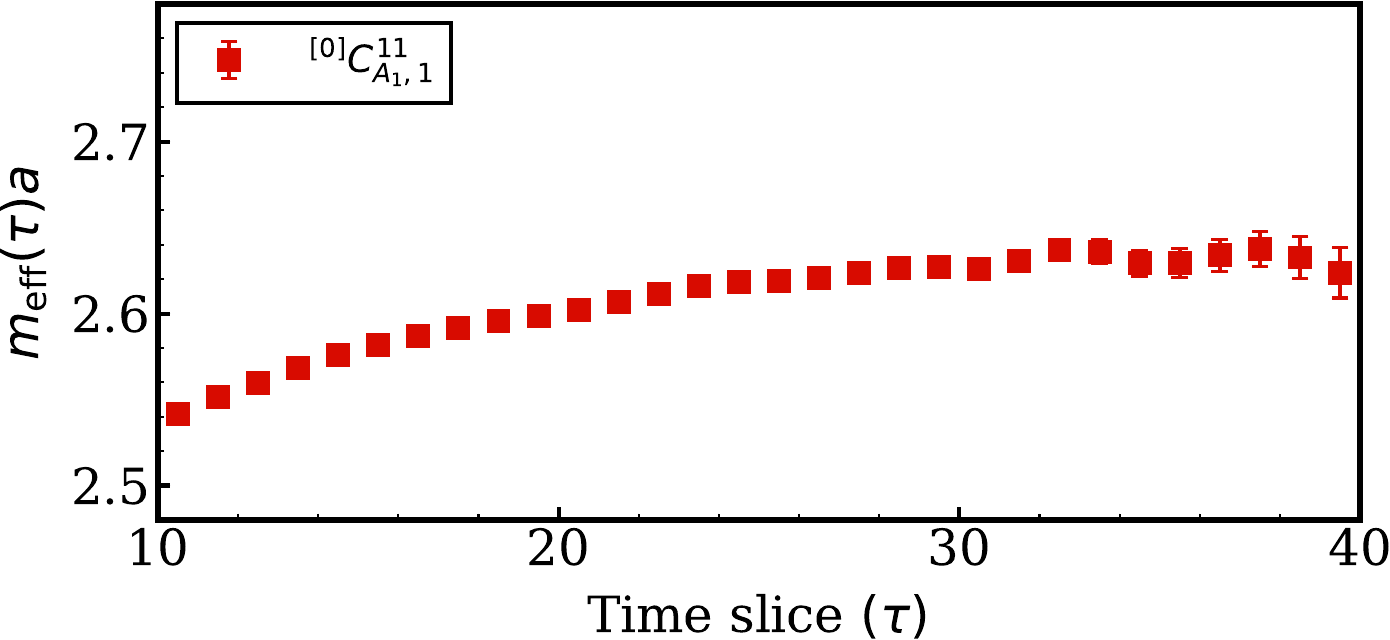}
\includegraphics[width=0.32\textwidth]{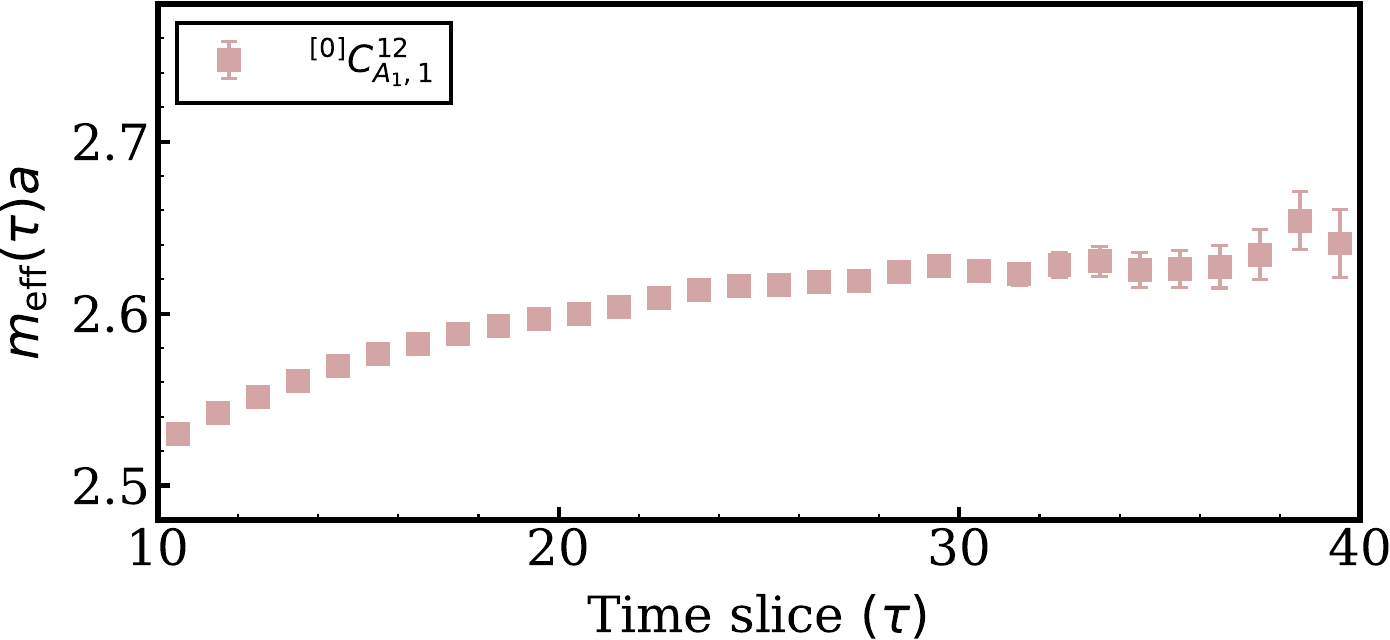}
\includegraphics[width=0.32\textwidth]{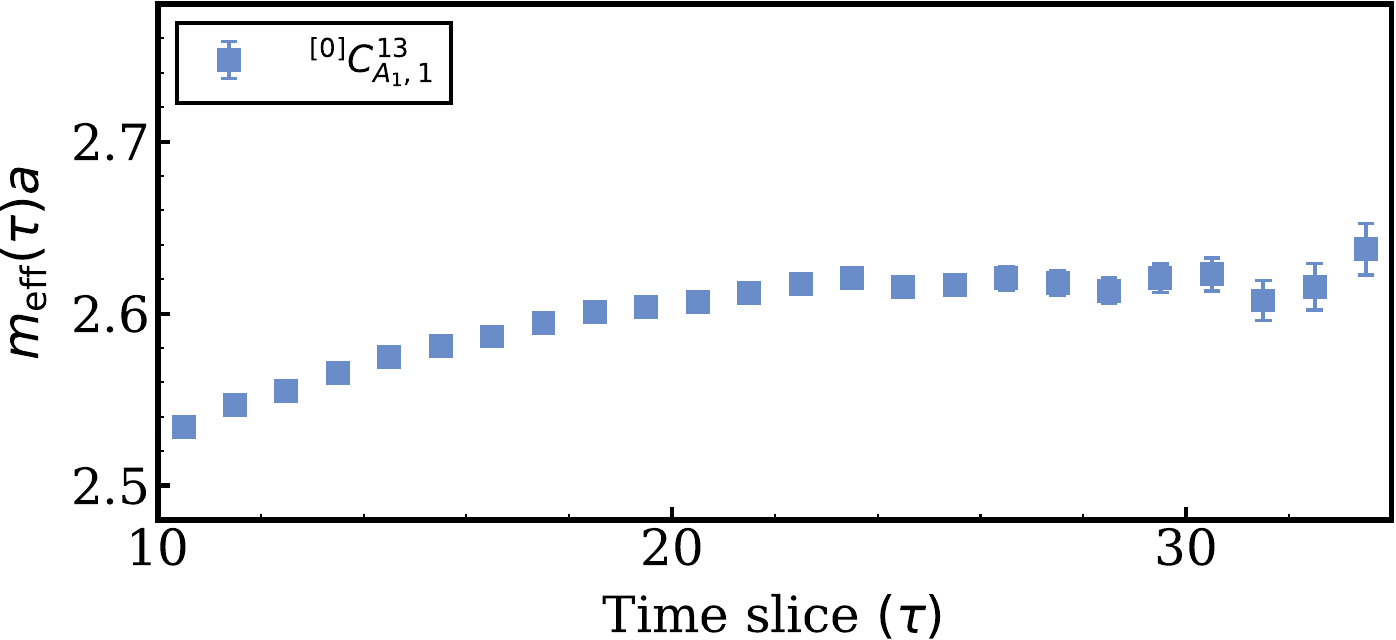}
\includegraphics[width=0.32\textwidth]{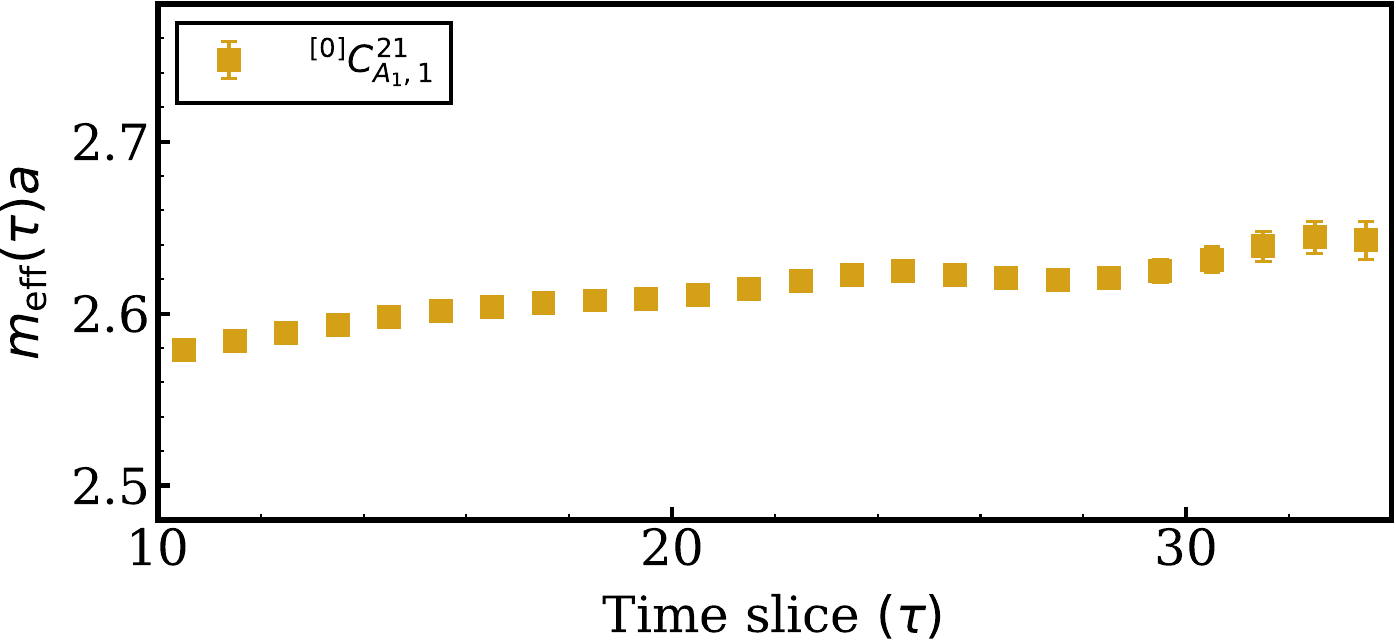}
\includegraphics[width=0.32\textwidth]{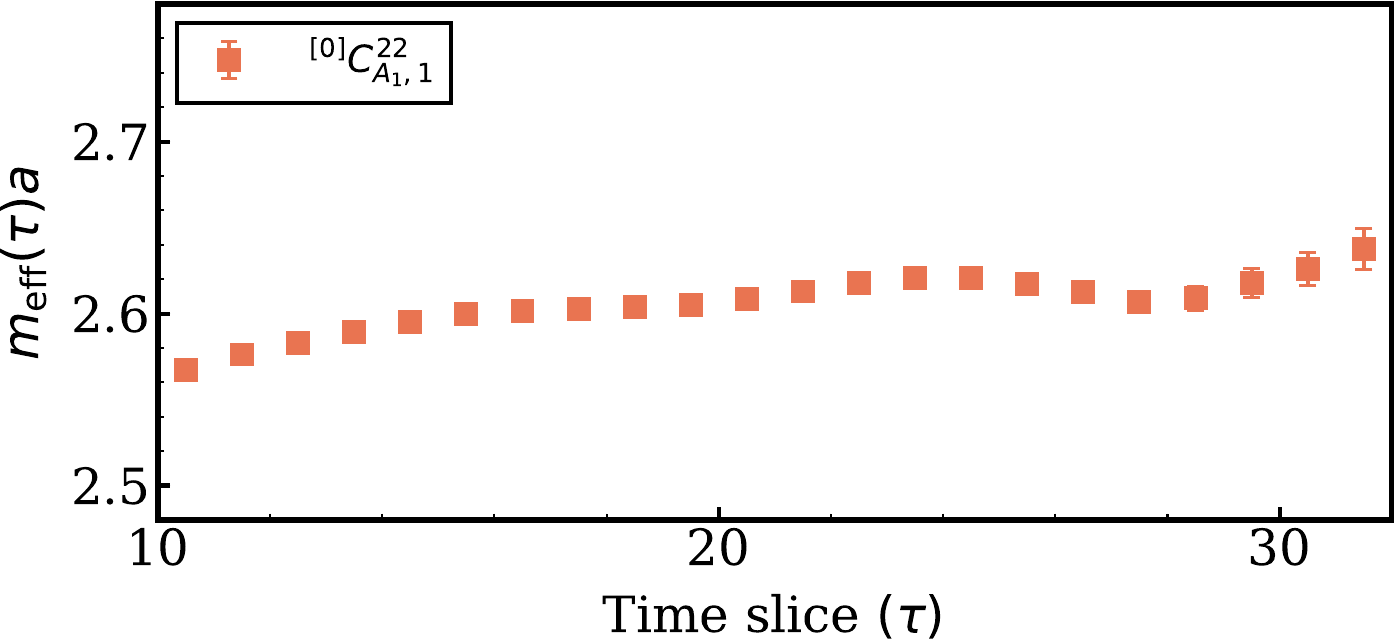}
\includegraphics[width=0.32\textwidth]{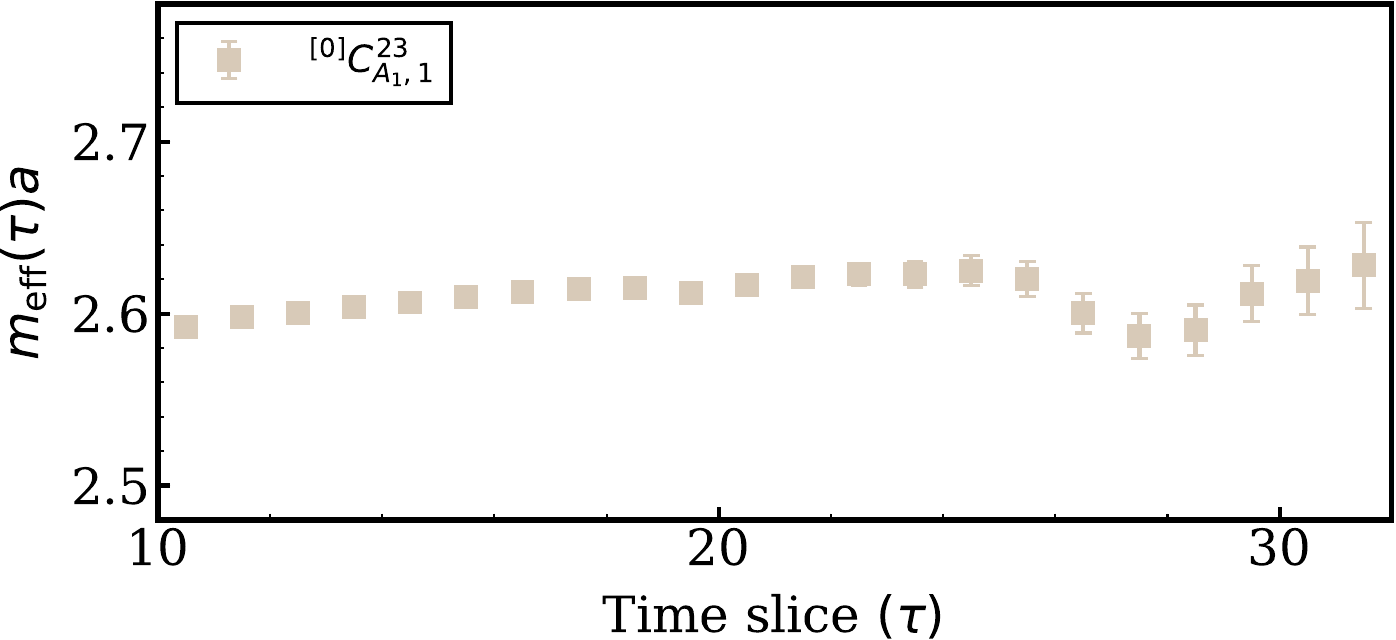}
\includegraphics[width=0.32\textwidth]{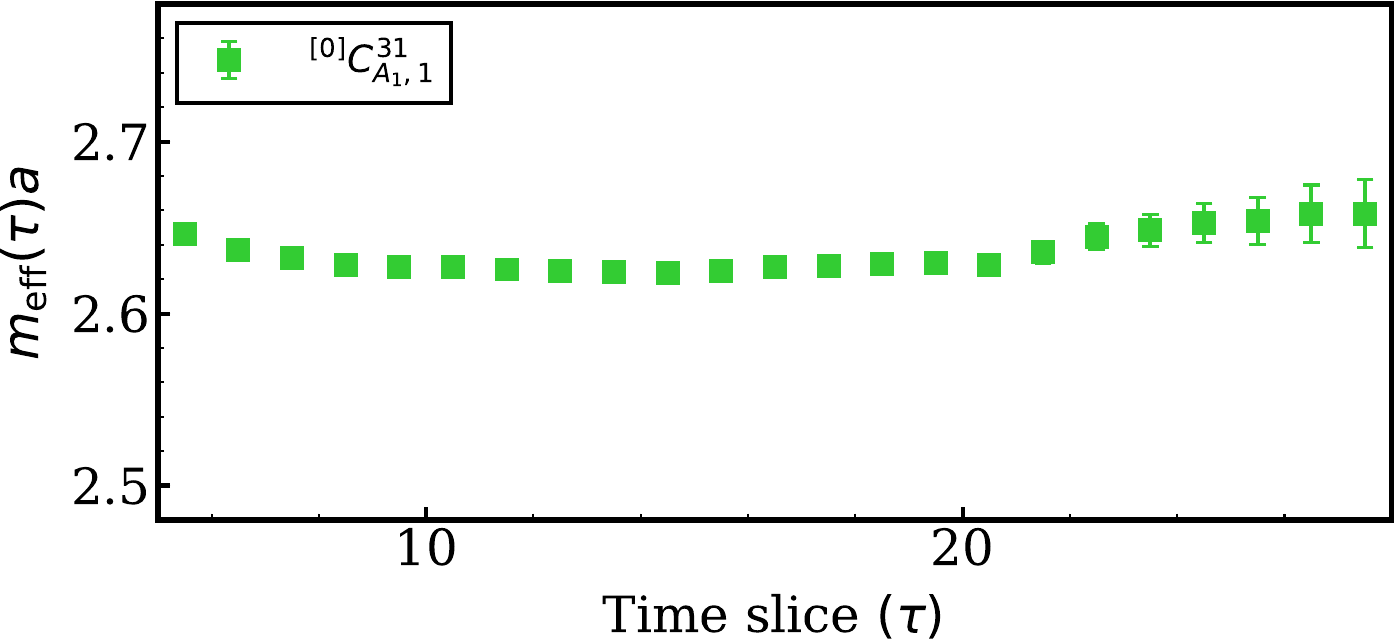}
\includegraphics[width=0.32\textwidth]{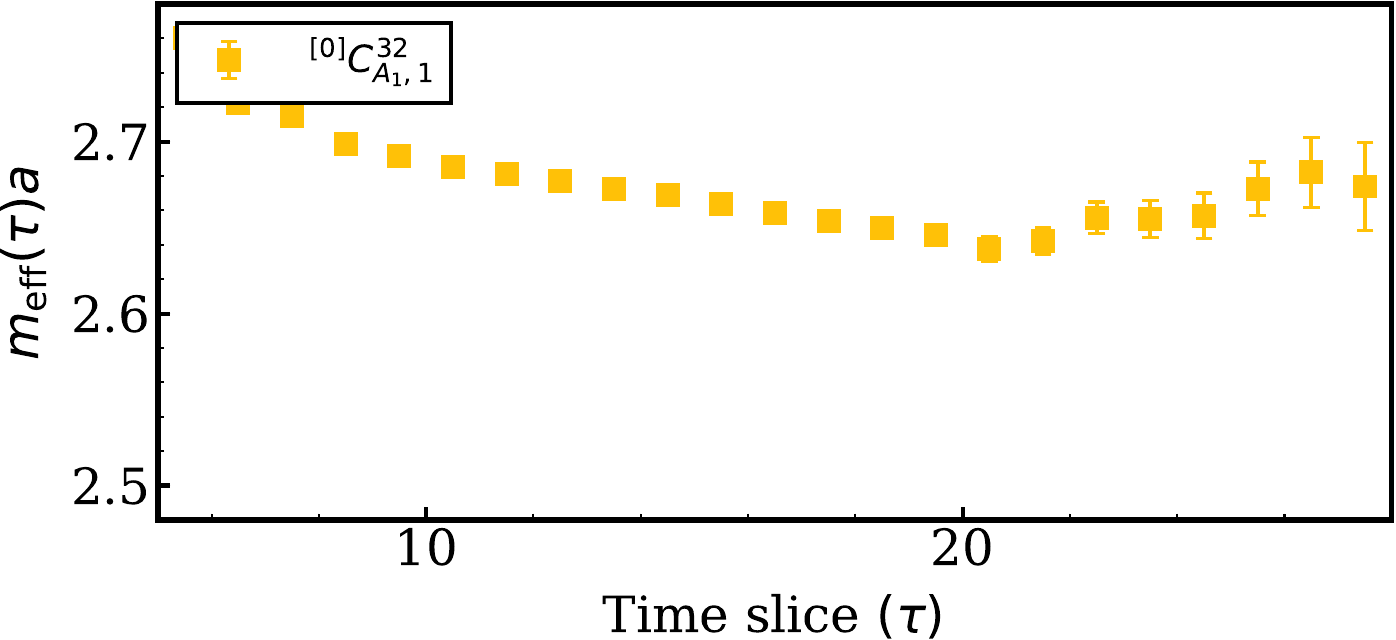}
\includegraphics[width=0.32\textwidth]{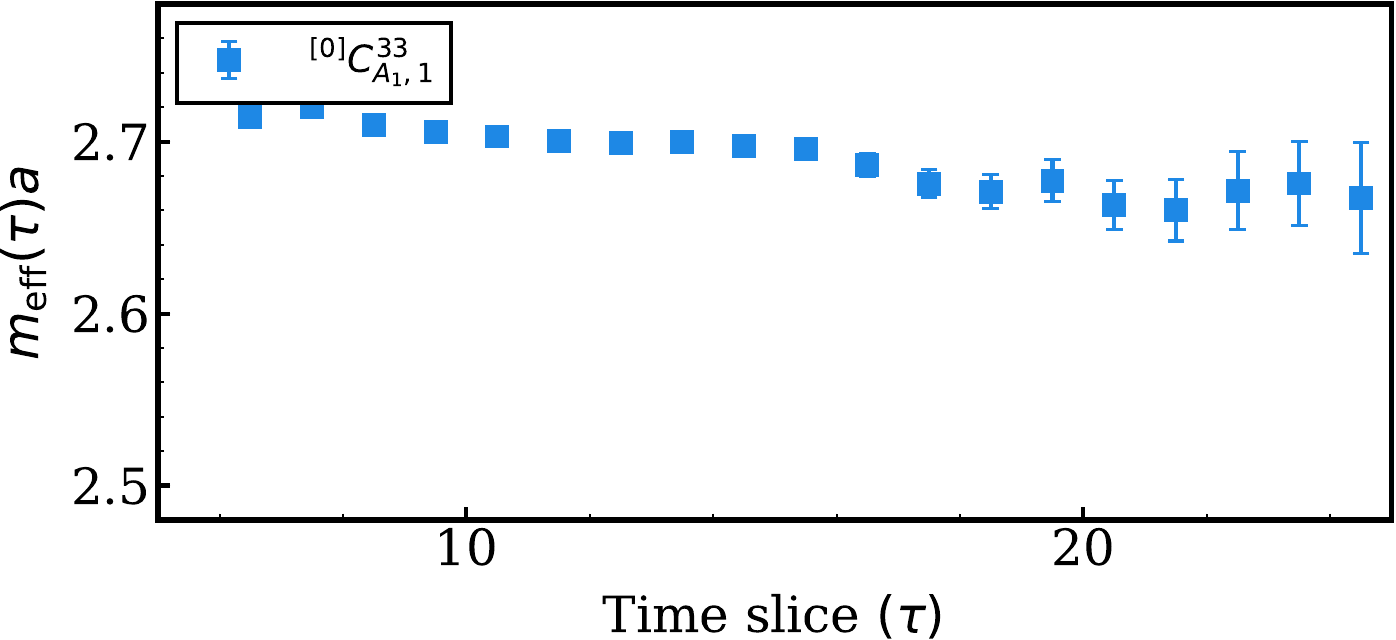}
\caption{\label{fig:app_op_dibar} Effective mass plots of the charmed scalar dibaryon $\mathcal{D}_{6c}$ correlation matrix elements [(Eq.~\eqref{eq:corr_dibar_s0})] on the $S_3$ lattice.}
\end{figure*}

\begin{figure*}[ht]
\includegraphics[width=0.32\textwidth]{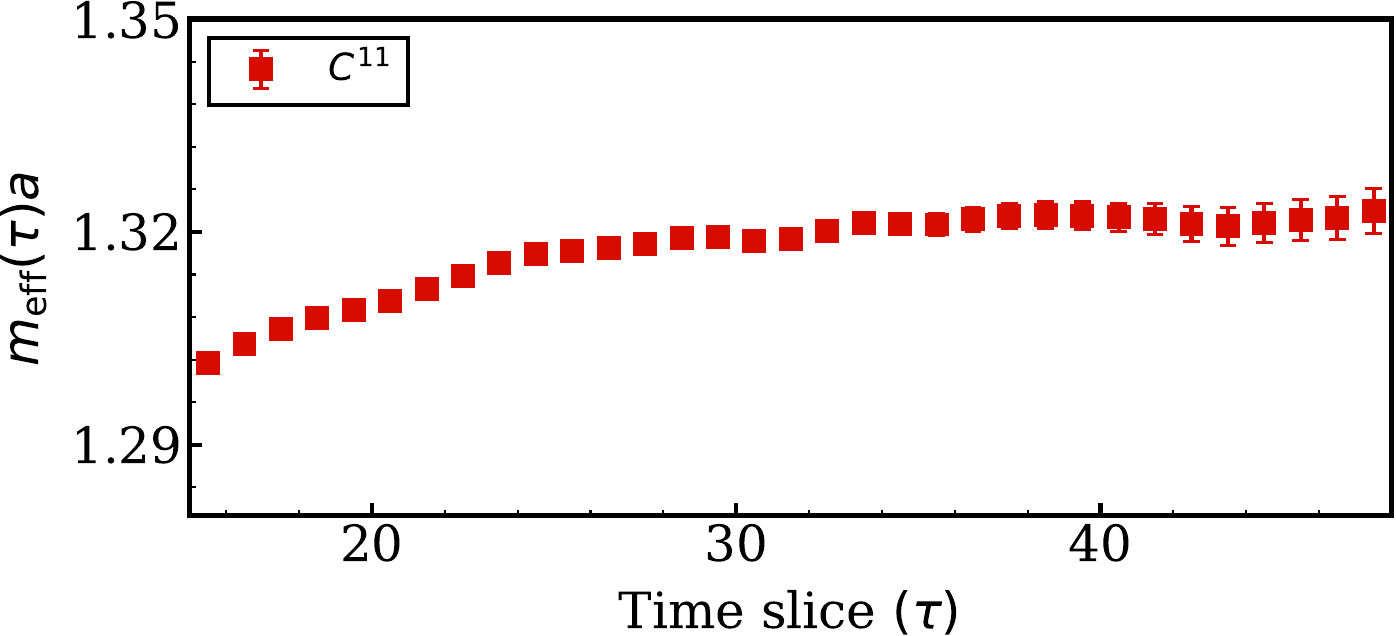}
\includegraphics[width=0.32\textwidth]{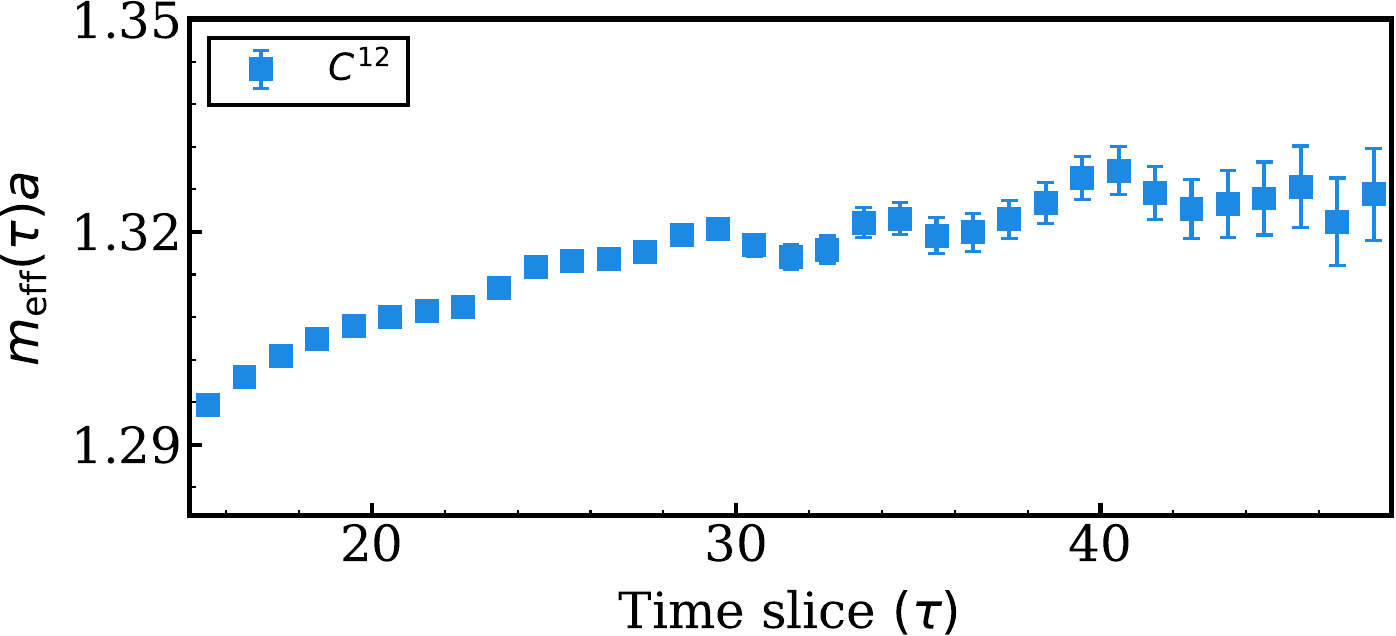}\\
\includegraphics[width=0.32\textwidth]{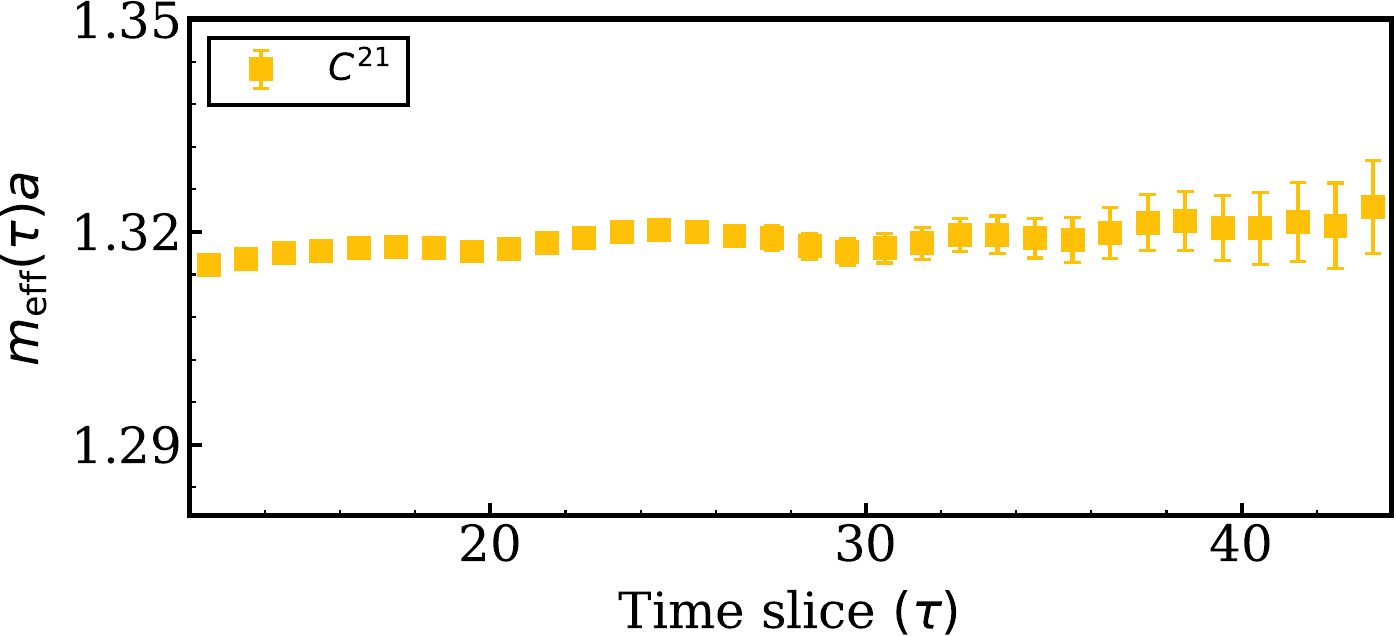}
\includegraphics[width=0.32\textwidth]{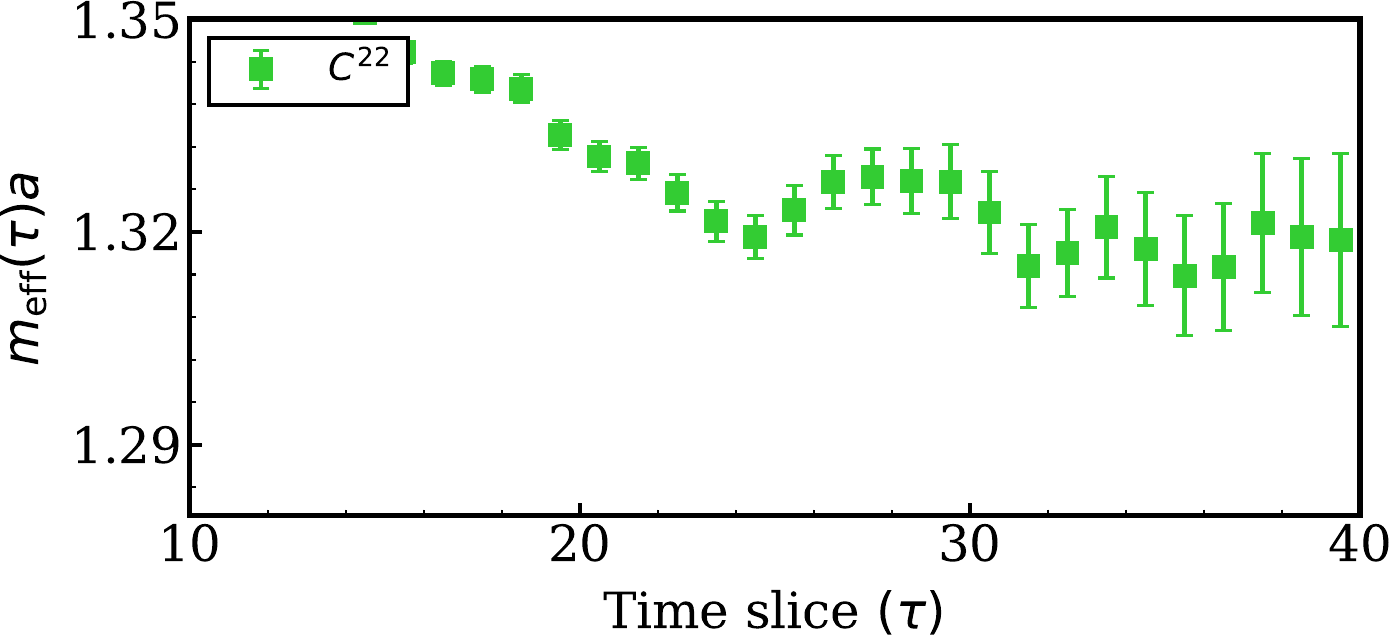}
\caption{\label{fig:app_op_bar} Effective mass plots of the $\Omega_{ccc}$ baryon correlation matrix elements (Eq.~\eqref{eq:corr_bar}) on the $S_3$ lattice.}
\end{figure*}

To complement the discussion in the main text, we present here a detailed analysis of effective masses obtained from individual correlation matrix elements, rather than those of the solutions of the generalized eigenvalue problem (GEVP). This allows for a clearer understanding of the relative contributions and signal quality from different operator structures. We focus on the $\Omega_{ccc}$ baryon and the corresponding spin-0 $\mathcal{D}_{6c}$ dibaryon channel. In Figures \ref{fig:app_op_dibar} and \ref{fig:app_op_bar}, we present the effective masses for the correlation matrix elements presented in Eqs. (\ref{eq:corr_dibar_s0}) and (\ref{eq:corr_bar}) for the charmed scalar dibaryons and the $\Omega_{ccc}$ baryons respectively, measured on the $S_3$ ensemble. The suffix superscripts in the legend labels $C_x^{ij}$ indicate the correlation matrix row ($i$) and column ($j$) entries, respectively.

The relativistic operators can be observed to mitigate the rising from below asymptotic behaviour of the correlation functions evaluated in the asymmetric setup we utilize. This behavior might be indicative of a more positive definite overlap factor, improving the fidelity of the signal. However, this increase in relativistic content also comes with enhanced statistical noise, making the individual correlators less stable. Nevertheless, when all these operator combinations are used together in the full correlation matrix, the generalized eigenvalue problem (GEVP) allows for a cleaner extraction of the ground state. The resulting mass estimates show better consistency and give us increased confidence in the extracted values.

\section{CG and Subduction coefficients}\label{app:cg}
Here, we present the Clebsch-Gordan (CG) and subduction coefficients used in constructing dibaryon operators. The CG coefficients ensure proper coupling of single baryon operators to form total angular momentum eigenstates. The subduction coefficients project these continuum spin states onto the relevant irreducible representations of the octahedral group in the finite volume. The list of CG coefficients for $S=0,2$ is given below:

\begin{itemize}
    \item $S=0,S_j=0$
\[
\begin{blockarray}{p{0.075\textwidth}|p{0.075\textwidth}p{0.075\textwidth}p{0.075\textwidth}p{0.075\textwidth}}
& 3/2 & 1/2 & -1/2 & -3/2  \\
\hline
\begin{block}{p{0.075\textwidth}|p{0.075\textwidth}p{0.075\textwidth}p{0.075\textwidth}p{0.075\textwidth}}
  3/2 & 0 & 0 & 0 & 1/2  \\
  1/2 & 0 & 0 & -1/2 & 0  \\
  -1/2 & 0 & 1/2 & 0 & 0  \\
  -3/2 & -1/2 & 0 & 0 & 0  \\
\end{block}
\end{blockarray}
 \]
 \item $S=2,S_j=0$
 \[
\begin{blockarray}{p{0.075\textwidth}|p{0.075\textwidth}p{0.075\textwidth}p{0.075\textwidth}p{0.075\textwidth}}
& 3/2 & 1/2 & -1/2 & -3/2  \\
\hline
\begin{block}{p{0.075\textwidth}|p{0.075\textwidth}p{0.075\textwidth}p{0.075\textwidth}p{0.075\textwidth}}
  3/2 & 0 & 0 & 0 & 1/2  \\
  1/2 & 0 & 0 & 1/2 & 0  \\
  -1/2 & 0 & -1/2 & 0 & 0  \\
  -3/2 & -1/2 & 0 & 0 & 0  \\
\end{block}
\end{blockarray}
 \]
\item $S=2,S_j=1$
 \[
\begin{blockarray}{p{0.075\textwidth}|p{0.075\textwidth}p{0.075\textwidth}p{0.075\textwidth}p{0.075\textwidth}}
& 3/2 & 1/2 & -1/2 & -3/2  \\
\hline
\begin{block}{p{0.075\textwidth}|p{0.075\textwidth}p{0.075\textwidth}p{0.075\textwidth}p{0.075\textwidth}}
  3/2 & 0 & 0 & $1/\sqrt{2}$ & 0  \\
  1/2 & 0 & 0 & 0 & 0  \\
  -1/2 & $-1/\sqrt{2}$ & 0 & 0 & 0  \\
  -3/2 & 0 & 0 & 0 & 0  \\
\end{block}
\end{blockarray}
 \]
\item $S=2,S_j=-1$ 
  \[
\begin{blockarray}{p{0.075\textwidth}|p{0.075\textwidth}p{0.075\textwidth}p{0.075\textwidth}p{0.075\textwidth}}
& 3/2 & 1/2 & -1/2 & -3/2  \\
\hline
\begin{block}{p{0.075\textwidth}|p{0.075\textwidth}p{0.075\textwidth}p{0.075\textwidth}p{0.075\textwidth}}
  3/2 & 0 & 0 & 0 & 0  \\
  1/2 & 0 & 0 & 0 & $1/\sqrt{2}$  \\
  -1/2 & 0 & 0 & 0 & 0  \\
  -3/2 & 0 & $-1/\sqrt{2}$ & 0 & 0  \\
\end{block}
\end{blockarray}
 \]
\item $S=2,S_j=2$ 
   \[
\begin{blockarray}{p{0.075\textwidth}|p{0.075\textwidth}p{0.075\textwidth}p{0.075\textwidth}p{0.075\textwidth}}
& 3/2 & 1/2 & -1/2 & -3/2  \\
\hline
\begin{block}{p{0.075\textwidth}|p{0.075\textwidth}p{0.075\textwidth}p{0.075\textwidth}p{0.075\textwidth}}
  3/2 & 0 & $1/\sqrt{2}$ & 0 & 0  \\
  1/2 & $-1/\sqrt{2}$ & 0 & 0 & 0  \\
  -1/2 & 0 & 0 & 0 & 0  \\
  -3/2 & 0 & 0 & 0 & 0  \\
\end{block}
\end{blockarray}
 \]
\item $S=2,S_j=-2$
 \[
\begin{blockarray}{p{0.075\textwidth}|p{0.075\textwidth}p{0.075\textwidth}p{0.075\textwidth}p{0.075\textwidth}}
& 3/2 & 1/2 & -1/2 & -3/2  \\
\hline
\begin{block}{p{0.075\textwidth}|p{0.075\textwidth}p{0.075\textwidth}p{0.075\textwidth}p{0.075\textwidth}}
  3/2 & 0 & 0 & 0 & 0  \\
  1/2 & 0 & 0 & 0 & 0  \\
  -1/2 & 0 & 0 & 0 & $1/\sqrt{2}$  \\
  -3/2 & 0 & 0 & $-1/\sqrt{2}$ & 0  \\
\end{block}
\end{blockarray}
 \]
\end{itemize}

We know that $S = 0$ continuum spin subduces only onto the one-dimensional $A_1$ irrep., hence:
\begin{equation}
    \mathcal{S}_{\text{$A_1$,1}}^{0,0} = 1, \mbox{~~and hence from Eq.~\eqref{eqn:subduction}~~} ^{[0]}\mathcal{O}^{a,b}_{\text{$A_1$,1}} = \mathcal{O}^{a,b}_{0}. \label{eq:a1_1}
\end{equation}
On the other hand, $S = 2$ continuum spin subduces onto the two-dimensional $E$ irrep. and onto the three-dimensional $T_2$ irrep of the Octahedral group. The subduction coefficients for $T_2$ irrep. are:
\[
\begin{blockarray}{p{0.03\textwidth}p{0.03\textwidth}|p{0.05\textwidth}p{0.05\textwidth}p{0.05\textwidth}p{0.05\textwidth}p{0.05\textwidth}}
&~~~$S_j$& 2 & 1 & 0 & -1& -2 \\
$\lambda$ && & & & &  \\
\hline
\begin{block}{p{0.03\textwidth}p{0.03\textwidth}|p{0.05\textwidth}p{0.05\textwidth}p{0.05\textwidth}p{0.05\textwidth}p{0.05\textwidth}}
  1& & 0 & 1 & 0 & 0 & 0\\
  2& & $1/\sqrt{2}$ & 0 & 0 & 0  & $-1/\sqrt{2}$\\
  3& & 0 & 0 & 0 & 1 & 0\\
\end{block}
\end{blockarray}
 \]
Using these subduction coefficients in Eq. \eqref{eqn:subduction} gives: 
\begin{align}
    ^{[2]}\mathcal{O}^{a,b}_{\text{$T_2$,1}} &= \mathcal{O}^{a,b}_1,\label{eq:t2_1}\\
    ^{[2]}\mathcal{O}^{a,b}_{\text{$T_2$,2}} &= \frac{1}{\sqrt{2}}\left(\mathcal{O}^{a,b}_2-\mathcal{O}^{a,b}_{-2}\right),\label{eq:t2_2}\\
    ^{[2]}\mathcal{O}^{a,b}_{\text{$T_2$,3}} &= \mathcal{O}^{a,b}_{-1}\label{eq:t2_3}.
\end{align}

Similarly, the subduction coefficients for $E$ irrep. are:
 \[
\begin{blockarray}{p{0.03\textwidth}p{0.03\textwidth}|p{0.05\textwidth}p{0.05\textwidth}p{0.05\textwidth}p{0.05\textwidth}p{0.05\textwidth}}
&~~~$S_j$& 2 & 1 & 0 & -1& -2 \\
$\lambda$&& & & & &  \\
\hline
\begin{block}{p{0.03\textwidth}p{0.03\textwidth}|p{0.05\textwidth}p{0.05\textwidth}p{0.05\textwidth}p{0.05\textwidth}p{0.05\textwidth}}
  1& & 0 & 0 & 1 & 0 & 0\\
  2& & $1/\sqrt{2}$ & 0 & 0 & 0  & $1/\sqrt{2}$\\
\end{block}
\end{blockarray}
 \]
 Again, using these subduction coefficients in Eq. \eqref{eqn:subduction} we get:
 \begin{align}
    ^{[2]}\mathcal{O}^{a,b}_{\text{$E$,1}} &= \mathcal{O}^{a,b}_0,\label{eq:e_1}\\
    ^{[2]}\mathcal{O}^{a,b}_{\text{$E$,2}} &= \frac{1}{\sqrt{2}}\left(\mathcal{O}^{a,b}_2+\mathcal{O}^{a,b}_{-2}\right)\label{eq:e_2}.
\end{align}
Using Eqs. \eqref{eq:a1_1}, \eqref{eq:t2_1}, \eqref{eq:t2_2}, \eqref{eq:t2_3}, \eqref{eq:e_1}, \eqref{eq:e_2} and the Clebsch-Gordan coefficients, we obtain the final dibaryon operators as presented in Eq. \eqref{eqn:ops}.
%%%%%%%%%%%%%%%%%%%%%%%%%%%%%%%%%%%%%%%%%%%%

\bibliography{Diomega}
\end{document}